\documentclass[apsrev4-1,twocolumn,footinbib,superscriptaddress,floatfix,bibliography]{revtex4-1}

\usepackage{times}
\usepackage{float}
\usepackage{multirow}
\usepackage[dvipsnames]{xcolor}
\colorlet{BLUE}{blue}
\usepackage{amsmath}
\usepackage{amsthm}
\usepackage{amssymb}
\usepackage{amsbsy}
\usepackage{enumitem}
\usepackage{wasysym}
\usepackage[english]{babel}
\usepackage[T1]{fontenc}
\usepackage[utf8]{inputenc} 
\usepackage{graphicx}
\usepackage{booktabs}
\usepackage{array}
\usepackage{makecell}
\usepackage{xurl}
\usepackage[colorlinks,bookmarks=false,citecolor=blue,linkcolor=red,urlcolor=blue]{hyperref}
\usepackage{ulem}
\usepackage{pstricks}
\usepackage{rotating}			       
\usepackage{tabularx,hhline}	
\usepackage[caption=false]{subfig}		
\newcolumntype{P}[1]{>{\centering\arraybackslash}p{#1}}

\usepackage{tikz}
\usepackage{etoolbox}

\newlength{\FigTwoAHeight}
\newlength{\FigTwoCTopGap}
\newcommand{\FigTwoALabel}{%
  \makebox[0pt][l]{%
    \raisebox{\dimexpr\FigTwoAHeight+0.35ex\relax}[0pt][0pt]{\normalsize\bfseries a}%
  }%
}

\newcommand{\fgiiType}[1]{%
  \makebox[3.40cm][l]{#1}%
}

\newcommand{\fgiiGeom}[1]{%
  \makebox[2.40cm][c]{%
    \raisebox{-0.43\height}{%
      \includegraphics[scale=0.52]{#1}%
    }%
  }%
}

\begin{document}

\preprint{APS/123-QED}

\title{Multipartite entanglement hidden in vector-chiral correlations of spin-\(1/2\) chains}

\author{Tokuro Shimokawa}

\email{tokuro.shimokawa@oist.jp}
\affiliation{Theory of Quantum Matter Unit, Okinawa Institute of Science and Technology Graduate University, Onna-son, Okinawa 904-0412, Japan}

\date{\today}

\begin{abstract}

Characterizing entanglement in quantum materials through experimentally
accessible observables remains a central challenge.
Quantum Fisher information associated with sums of one-site spin
operators provides one route by linking entanglement to measurable
magnetic response functions.
Extending this approach to local composite observables requires
new entanglement bounds.
Here we establish a rigorous bound for the total nearest-neighbour
vector chirality of periodic spin-$1/2$ chains with an even number
of sites.
The quantum Fisher information per spin cannot exceed $k+1$ for any
$k$-producible state, so exceeding this bound certifies entanglement
depth of at least $k+1$.
Combining infinite density-matrix renormalization group and thermal
pure quantum-state calculations, we demonstrate multipartite
entanglement encoded in vector-chiral correlations at zero and finite
temperatures in an extended $J_1$--$J_2$ XXZ chain.
The spin-current coupling between vector chirality and electric
polarization offers, in principle, access to this witness through
dielectric loss spectroscopy in multiferroic magnets, extending
entanglement diagnostics beyond dipolar spin correlations.

\end{abstract}

\maketitle

\section{Introduction}

How can entanglement in quantum materials be characterized and
measured experimentally?
In principle, complete knowledge of the many-body density matrix
contains all the information required to characterize a quantum
state, including its entanglement properties.
In bulk quantum materials, however, reconstructing the full many-body
density matrix through quantum-state tomography is highly challenging.
Experimental studies therefore often rely on a restricted set of
measurable observables from which entanglement can be inferred or
certified.
Several experimentally accessible entanglement diagnostics have been
developed for this purpose~\cite{Guhne2009, Laurell2025}.
Such approaches generally provide information about particular aspects
of entanglement.
Developing complementary diagnostics that access different aspects
of entanglement while remaining experimentally measurable is therefore
an important challenge.

Quantum Fisher information (QFI) provides an important route towards
this goal.
It quantifies the sensitivity of a quantum state, including a mixed
state, to a unitary rotation generated by a chosen observable~\cite{Toth2014}.
In quantum magnets, QFI-based entanglement certification has
predominantly used collective spin generators constructed as linear
sums of one-site operators, such as
$H_{\rm lin}=\sum_i \mathbf n_i\cdot\mathbf S_i$.
For these generators, $k$-producible upper bounds on the QFI were independently established by Hyllus {\it et al}. and Tóth, allowing
the QFI to place a lower bound on entanglement depth, thereby
certifying multipartite entanglement~\cite{Hyllus2012,Toth2012}.
In thermal equilibrium, QFI can be related to experimentally accessible linear-response functions, including magnetic susceptibilities and dynamical structure factors
~\cite{Hauke2016,Shitara2016}.
This connection has enabled both experimental and theoretical
studies of entanglement through spectroscopic response functions
in quantum materials
~\cite{Laurell2021,Scheie2021,Pratt2022,Menon2023,Scheie2024,
Laurell2025,Zhang2025,David2025,Wu2025,Sid2025, Tokuro2025}.
We refer to witnesses built from such one-site linear spin generators
as spin-based QFI witnesses.

However, spin-based QFI provides a sufficient, but not necessary,
criterion for multipartite entanglement: failure to exceed a
$k$-producible bound does not imply the absence of multipartite
entanglement.
For example, spin-based QFI did not certify genuine multipartite
entanglement in a study of the Kitaev honeycomb model
~\cite{Lambert2020}, whereas subsequent reduced-density-matrix-based
studies using PPT-based multipartite-entanglement measures have
revealed spatially structured multipartite entanglement at zero and
finite temperatures in Kitaev- and Kagome-lattice spin-liquid models
~\cite{Lyu2026,Sid2025v2}.
These results motivate extending experimentally accessible
entanglement diagnostics beyond the dipolar spin correlations
probed by spin-based QFI.


In this work, we pursue one such route by constructing QFI generators
from sums of local composite observables, thereby probing higher-order
spin correlations.
For such a QFI to certify entanglement, a corresponding
$k$-producible upper bound must be established.
The central difficulty is that a local composite observable can act
on spins belonging to different blocks of a $k$-producible
partition, whose blocks each contain at most $k$ spins.
Unlike a sum of one-site spin operators, the generator therefore
need not admit a decomposition into contributions acting entirely
within individual blocks.
Establishing a bound that accounts for these cross-block terms is
the first step towards extending QFI-based entanglement certification
to local composite observables.

Here, as a concrete example of this broader strategy, we construct
such a bound for a local composite generator based on
nearest-neighbour vector chirality in spin-$1/2$ chains, while
retaining a route to experimental access.
Vector chirality characterizes the handedness of noncollinear spin
configurations and has played an important role across a broad range
of classical and quantum frustrated magnets
~\cite{Miyashita1984,Kawamura1984,Kawamura1998,Nersesyan1998,
Hikihara2001,Hikihara2008,Furukawa2010}.
As a two-spin observable defined on bonds, its bond--bond correlations
probe a channel distinct from the dipolar correlations entering
spin-based QFI.
Moreover, vector chirality is coupled to electric polarization in
multiferroic magnets, as described by the spin-current mechanism and
related phenomenological theories and demonstrated experimentally in
quantum spin-chain compounds
~\cite{Katsura2005,Katsura2007,Mostovoy2006,Seki2008}, providing a
connection between the response of this composite observable and
dielectric spectroscopy.

Specifically, we consider the total vector-chirality operator
\begin{equation}
K_{\rm tot}
=
\sum_i \kappa_i,
\qquad
\kappa_i =
(\mathbf S_i\times \mathbf S_{i+1})^z
=
S_i^x S_{i+1}^y
-
S_i^y S_{i+1}^x .
\label{eq:Ktot-def}
\end{equation}
For an even number $N\ge4$ of sites and periodic boundary conditions,
we derive a rigorous chirality-QFI density bound for all
$k$-producible states:
\begin{equation}
\frac{F_Q[\rho^{(k)},K_{\rm tot}]}{N}
\le
k+1 .
\label{eq:kplusone-bound-simple}
\end{equation}
Here $F_Q[\rho,A]$ denotes the QFI of a state $\rho$ with respect
to a Hermitian generator $A$. For the spectral decomposition
\(\rho=\sum_m p_m|m\rangle\langle m|\),
the QFI is given by
\begin{equation}
F_Q[\rho,A]
=
2
\sum_{\substack{m,n\\p_m+p_n>0}}
\frac{(p_m-p_n)^2}{p_m+p_n}
\left|
\langle m|A|n\rangle
\right|^2,
\label{eq:QFI-def-intro}
\end{equation}
where \(p_m\) and \(|m\rangle\) denote the eigenvalues and
eigenstates of \(\rho\), respectively~\cite{Toth2014,Liu2020}.
The notation $\rho^{(k)}$ denotes a $k$-producible state, namely a
state with entanglement depth at most $k$; its formal definition is
given in Sec.~II.
Importantly, this bound is Hamiltonian-independent within this class of
even periodic spin-$1/2$ chains, provided that the QFI generator is the
total nearest-neighbour vector chirality \(K_{\rm tot}\) defined in
Eq.~\eqref{eq:Ktot-def}.
Consequently, a state whose chirality-QFI density exceeds the
corresponding threshold must have entanglement depth of at least
$k+1$.

\begin{figure}[t]
\centering
\includegraphics[width=0.45\textwidth]{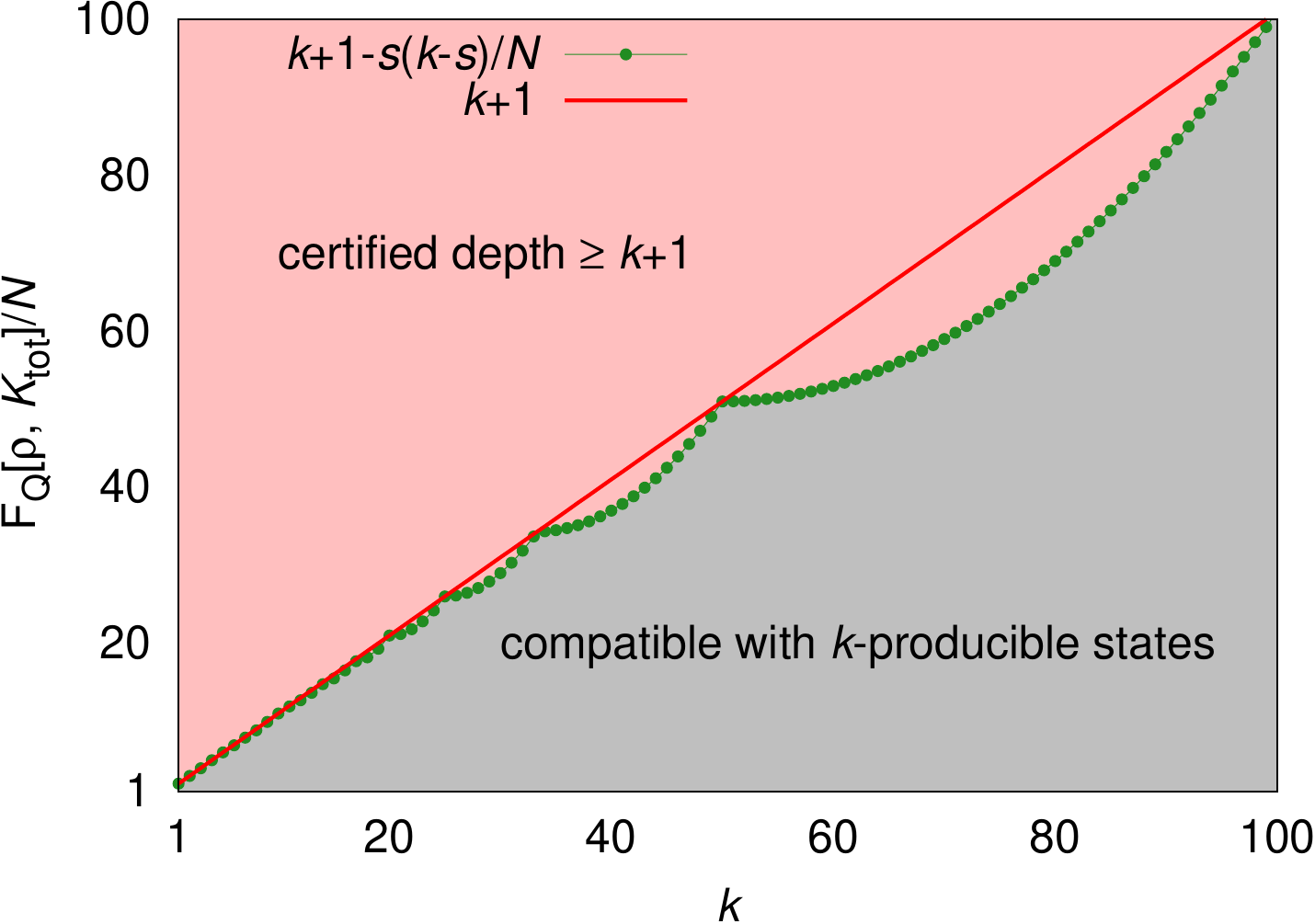}

\caption{
Multipartite-entanglement threshold for the vector-chirality QFI.
The plot shows the rigorous $k$-producible upper bound for the vector-chirality QFI density at $N=100$.
The green curve shows the finite-size threshold $k+1-s(k-s)/N$, with $N=qk+s$, while the red line shows the simplified bound $k+1$.
QFI densities above the green threshold certify entanglement depth at least $k+1$; the grey region is compatible with $k$-producible states.
}

\label{fig:chirality-threshold}
\end{figure}

We then apply this witness to a physically motivated extended
spin-$1/2$ $J_1$--$J_2$ XXZ chain introduced for one-dimensional
multiferroics with vector-chiral correlations
~\cite{Nersesyan1998,Hikihara2001,Onoda2007,Furukawa2008,
Hikihara2008,Furukawa2010,Sato2011,Furukawa2012}.
Using TeNPy-based iDMRG at zero temperature and thermal pure
quantum-state calculations
~\cite{Imada1986,Hams2000,Sugiura2013,Ikeuchi2015,Endo2018,Tokuro2025}
at finite temperatures, we evaluate the chirality QFI and the
corresponding certified entanglement depth.
We find that vector-chiral correlations host multipartite
entanglement at both zero and finite temperatures, demonstrating
the applicability of the composite-observable entanglement witness
to a concrete model relevant to frustrated quantum magnets.

Finally, we connect this entanglement certification to experiment.
Through the spin-current coupling between vector chirality and
electric polarization
~\cite{Katsura2005,Katsura2007,Mostovoy2006,Seki2008},
the dynamical susceptibility entering the chirality QFI can, in
principle, be accessed through the vector-chirality contribution to
the dielectric loss spectrum $\varepsilon''(\omega)$
~\cite{Furukawa2008,Furukawa2010,Grams2022}.
This provides a route to certify multipartite entanglement encoded
in vector-chiral correlations from dielectric spectroscopy in
candidate quasi-one-dimensional multiferroic magnets such as
LiCuVO$_4$ and related frustrated chain compounds
~\cite{Schrettle2008,Enderle2010,Yasui2013,Reynolds2019}.




\section{Chirality-QFI entanglement criterion}

A pure state is \(k\)-producible if there exists a partition
\(\Gamma\) of the \(N\) sites into blocks \(u\) such that
\begin{equation}
|\Psi^{(k)}\rangle
=
\bigotimes_{u\in\Gamma}|\psi_u\rangle,
\qquad
n_u\le k,
\label{eq:kprod-pure}
\end{equation}
where \(n_u\) denotes the number of sites in block \(u\), and the
blocks need not be spatially contiguous.
The definition is inclusive: any state that is \(k'\)-producible
for some \(k'\le k\) is also \(k\)-producible.
Thus, no assumption is made that the state within an allowed block
is genuinely \(n_u\)-partite entangled; if it further factorizes,
the state simply admits a finer \(k\)-producible partition.

A mixed state is \(k\)-producible if it can be written as a
convex mixture of such pure states,
\begin{equation}
\rho^{(k)}
=
\sum_{\lambda}
p_{\lambda}
|\Psi^{(k)}_{\lambda}\rangle
\langle\Psi^{(k)}_{\lambda}|,
\qquad
p_{\lambda}\ge0,
\qquad
\sum_{\lambda}p_{\lambda}=1 ,
\label{eq:kprod-mixed}
\end{equation}
where the block partition may depend on \(\lambda\).
A state that is \(k\)-producible has entanglement depth at most
\(k\).

The central result of this section is the finite-size refinement
of Eq.~\eqref{eq:kplusone-bound-simple}.
For a spin-\(1/2\) chain with an even number \(N\ge4\) of sites
and periodic boundary conditions, and for \(1\le k\le N\), write
\begin{equation}
N=qk+s,
\qquad
0\le s<k .
\label{eq:N-qks}
\end{equation}
Then every \(k\)-producible state satisfies
\begin{equation}
\frac{F_Q[\rho^{(k)},K_{\rm tot}]}{N}
\le
\frac{qk^2+s^2+N}{N}
=
k+1-\frac{s(k-s)}{N}.
\label{eq:kplusone-bound-finite}
\end{equation}
Since \(s(k-s)\ge0\), this finite-size result immediately implies
the simpler bound
\(F_Q[\rho^{(k)},K_{\rm tot}]/N\le k+1\)
stated in Eq.~\eqref{eq:kplusone-bound-simple}.
A QFI density exceeding the right-hand side of
Eq.~\eqref{eq:kplusone-bound-finite} therefore certifies
entanglement depth at least \(k+1\).
Figure~\ref{fig:chirality-threshold} illustrates the finite-size
\(k\)-producible threshold for \(N=100\), together with the simplified
bound \(k+1\).

\begin{figure*}[t]
\centering

\begin{minipage}[t]{0.35\textwidth}
\vspace*{8mm}
\centering

\raisebox{1.2ex}[0pt][0pt]{\FigTwoALabel}
\includegraphics[height=\FigTwoAHeight]{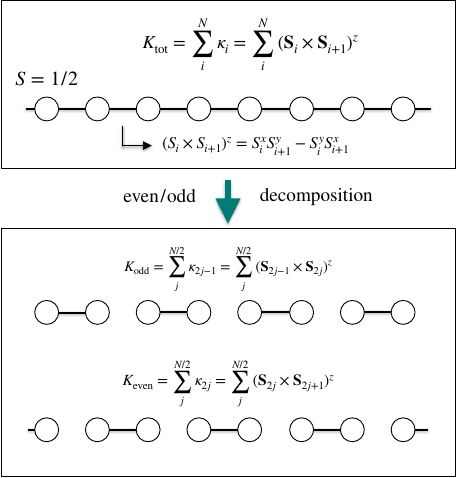}

\end{minipage}%
\hspace{0.005\textwidth}%
\begin{minipage}[t]{0.64\textwidth}
\vspace*{7.9mm}
\centering

\scriptsize
\setlength{\tabcolsep}{0pt}
\renewcommand{\arraystretch}{1.14}

\makebox[\linewidth][l]{%
\makebox[0pt][l]{%
  \hspace*{-0.8mm}%
  \raisebox{2.0ex}[0pt][0pt]{{\normalsize\bfseries c}}%
}%
\begin{tabular}[t]{@{} l l l l @{}}
  \toprule

  \makebox[0.30\linewidth][l]{\textbf{bond type}}
  &
  \makebox[0.22\linewidth][c]{\textbf{block geometry}}
  &
  \makebox[0.30\linewidth][c]{\textbf{term bound}}
  &
  \makebox[0.18\linewidth][c]{\textbf{contribution to \(A_u\)}}
  \\
  \midrule

  \makebox[0.30\linewidth][l]{\fgiiType{internal}}
  &
  \makebox[0.22\linewidth][c]{\fgiiGeom{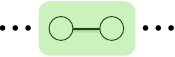}}
  &
  \makebox[0.30\linewidth][c]{%
    \(\mathrm{Var}(\kappa_e^{\rm in})\le 1/4\)
  }
  &
  \makebox[0.18\linewidth][c]{\(r_u/4\)}
  \\[0.65ex]

  \makebox[0.30\linewidth][l]{\fgiiType{cross}}
  &
  \makebox[0.22\linewidth][c]{\fgiiGeom{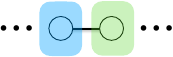}}
  &
  \makebox[0.30\linewidth][c]{%
    \(\mathrm{Var}(\kappa_e^{\rm cr})\le 1/4\)
  }
  &
  \makebox[0.18\linewidth][c]{\(c_u/8\)}
  \\
  \bottomrule
\end{tabular}
}

\vspace{\FigTwoCTopGap}
\vspace{1.55mm}

\scriptsize
\setlength{\tabcolsep}{0pt}
\renewcommand{\arraystretch}{1.14}

\makebox[\linewidth][l]{%
\makebox[0pt][l]{%
  \hspace*{-0.8mm}%
  \raisebox{2.0ex}[0pt][0pt]{{\normalsize\bfseries d}}%
}%
\begin{tabular}[t]{@{} l l l l @{}}
  \toprule

  \makebox[0.30\linewidth][l]{\textbf{bond-pair type}}
  &
  \makebox[0.22\linewidth][c]{\textbf{block geometry}}
  &
  \makebox[0.30\linewidth][c]{\textbf{term bound}}
  &
  \makebox[0.18\linewidth][c]{\textbf{contribution to \(A_u\)}}
  \\
  \midrule

  \makebox[0.30\linewidth][l]{\fgiiType{internal--internal}}
  &
  \makebox[0.22\linewidth][c]{\fgiiGeom{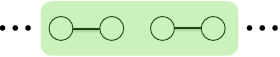}}
  &
  \makebox[0.30\linewidth][c]{%
    \(2|\mathrm{Cov}(\kappa_e^{\rm in},\kappa_f^{\rm in})|
    \!\le\!1/2\)
  }
  &
  \makebox[0.18\linewidth][c]{%
    \((1/2)\binom{r_u}{2}\)
  }
  \\[0.65ex]

  \makebox[0.30\linewidth][l]{\fgiiType{internal--cross}}
  &
  \makebox[0.22\linewidth][c]{\fgiiGeom{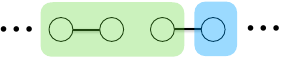}}
  &
  \makebox[0.30\linewidth][c]{%
    \(2|\mathrm{Cov}(\kappa_e^{\rm in},\kappa_f^{\rm cr})|
    \!\le\!1/4\)
  }
  &
  \makebox[0.18\linewidth][c]{%
    \((1/4)r_uc_u\)
  }
  \\[0.65ex]

  \makebox[0.30\linewidth][l]{%
    \fgiiType{cross--cross: one shared block}%
  }
  &
  \makebox[0.22\linewidth][c]{\fgiiGeom{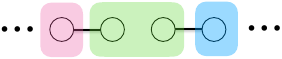}}
  &
  \makebox[0.30\linewidth][c]{%
    \(2|\mathrm{Cov}(\kappa_e^{\rm cr},\kappa_f^{\rm cr})|
    \!\le\!1/8\)
  }
  &
  \makebox[0.18\linewidth][c]{%
    \raisebox{-1.35ex}[0pt][0pt]{%
      \((1/8)\binom{c_u}{2}\)%
    }%
  }
  \\[0.65ex]

  \makebox[0.30\linewidth][l]{%
    \fgiiType{cross--cross: same two blocks}%
  }
  &
  \makebox[0.22\linewidth][c]{\fgiiGeom{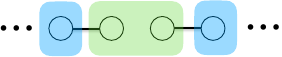}}
  &
  \makebox[0.30\linewidth][c]{%
    \(2|\mathrm{Cov}(\kappa_e^{\rm cr},\kappa_f^{\rm cr})|
    \!\le\!1/4\)
  }
  &
  \makebox[0.18\linewidth][c]{}
  \\
  \bottomrule
\end{tabular}
}

\end{minipage}

\vspace{4.0mm}

\noindent
\makebox[\textwidth][l]{%
  \hspace*{0.030\textwidth}\textbf{b}%
}

\vspace{-0.5ex}

\makebox[\textwidth][c]{%
  \includegraphics[width=0.76\textwidth]{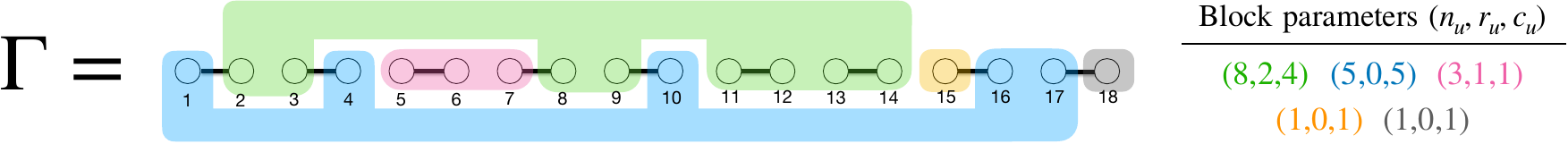}%
}

\caption{
Structure of the chirality-QFI bound.
\textbf{a}, Even--odd decomposition of the total vector-chirality
operator \(K_{\rm tot}\) into \(K_{\rm odd}\) and \(K_{\rm even}\).
Each component is a sum of mutually non-overlapping nearest-neighbour
chirality bonds.
\textbf{b}, Example of a block partition compatible with a
\(k=9\)-producible pure state for \(N=18\), shown relative to the
odd matching.
Each coloured region denotes one block in the chosen product
decomposition and contains at most \(k\) sites; sites belonging to
the same block need not be spatially contiguous.
No genuine multipartite entanglement within an individual block is
assumed: the state of a block may itself factorize into smaller
subsets.
Black solid bonds denote the chirality bonds belonging to
\(K_{\rm odd}\).
A matching bond whose two endpoints belong to the same block is an
internal bond, whereas a bond connecting two different blocks is a
cross bond.
The block parameters shown on the right are
\((n_u,r_u,c_u)\), where \(n_u\) is the number of sites in block
\(u\), \(r_u\) is the number of internal matching bonds in that
block, and \(c_u\) is the number of cross matching bonds incident on
it.
In \textbf{c} and \textbf{d}, the green-coloured block is denoted by
\(u\), and the block geometries shown are drawn with respect to this
choice.
\textbf{c}, Single-bond variance contributions for one matching
component.
The term bound is
\(\mathrm{Var}(\kappa_e)\le 1/4\) for both internal and cross bonds.
The contribution-to-\(A_u\) column shows the corresponding
block-wise contributions \(r_u/4\) and \(c_u/8\).
\textbf{d}, Block-wise classification of bond-pair covariance
geometries entering the variance of one matching component.
The term bound column gives the full covariance contribution
\(2|\mathrm{Cov}(\kappa_e,\kappa_f)|\) appearing in the variance
expansion.
The contribution-to-\(A_u\) column shows the corresponding
block-wise contributions to \(A_u\).
Internal--internal pairs give
\((1/2)\binom{r_u}{2}\), whereas internal--cross pairs give
\((1/4)r_uc_u\).
The two cross--cross geometries together exhaust the
\(\binom{c_u}{2}\) unordered pairs of cross bonds incident on block
\(u\), and their combined contribution to \(A_u\) is therefore
\((1/8)\binom{c_u}{2}\).
For a same-two-block cross--cross pair, the same pair is also counted
for the other shared block, so the two \(1/8\) allocations together
cover the full \(1/4\) term bound.
Identical colours in the same row indicate the same block, which need
not be spatially contiguous.
}
\label{fig:chirality-bound-structure}
\end{figure*}

The proof proceeds in six steps.
Step 1 uses the convexity of the QFI to reduce the problem to pure
\(k\)-producible states, for which the QFI is four times the
variance of the generator.
Step 2 decomposes \(K_{\rm tot}\) into two complete matchings.
Step 3 expands the variance within one matching into
single-bond variances and pairwise covariances, which are then
bounded according to their block geometries.
Step 4 collects these local bounds into block-wise contributions.
Step 5 maximizes the resulting expression over all pure
\(k\)-producible states, including all admissible block
partitions, under the \(k\)-producibility constraint.
Step 6 recombines the odd and even matching bounds using
the triangle inequality for the square root of the QFI.

\vspace{1.0ex}
\subsection*{Step 1: Reduction to pure \(k\)-producible states}

By convexity of the QFI and Eq.~\eqref{eq:kprod-mixed},
\begin{equation}
F_Q[\rho^{(k)},K_{\rm tot}]
\le
\sum_{\lambda}
p_{\lambda}
F_Q[
|\Psi^{(k)}_{\lambda}\rangle,
K_{\rm tot}
].
\label{eq:QFI-convex-main}
\end{equation}
It is therefore sufficient to establish the same uniform upper
bound for every pure \(k\)-producible state.
Let \(\mathcal P_k\) denote the set of all partitions
\(\Gamma\) of the \(N\) sites into blocks satisfying
\(n_u\le k\).
For Steps 2--5, we fix an arbitrary pure \(k\)-producible state
\(|\Psi^{(k)}\rangle\) together with one admissible partition
\(\Gamma\in\mathcal P_k\) for which
Eq.~\eqref{eq:kprod-pure} holds.
For a pure \(k\)-producible state,
\begin{equation}
F_Q[|\Psi^{(k)}\rangle,K_{\rm tot}]
=
4\operatorname{Var}_{\Psi^{(k)}}(K_{\rm tot}),
\label{eq:pure-QFI-var-main}
\end{equation}
so the problem reduces to bounding the variance of
\(K_{\rm tot}\) over pure \(k\)-producible block-product states.
In Steps 2--4, all expectation values, variances, and
covariances are evaluated in this fixed but otherwise arbitrary
pure \(k\)-producible state \(|\Psi^{(k)}\rangle\), and this
state dependence is displayed explicitly in the notation.

\vspace{1.0ex}
\subsection*{Step 2: Matching decomposition}

The even--odd decomposition is
\begin{align}
K_{\rm tot}
&=
K_{\rm odd}+K_{\rm even},
\label{eq:Ktot-odd-even}
\\
K_{\rm odd}
&=
\sum_{j=1}^{N/2}\kappa_{2j-1},
\qquad
K_{\rm even}
=
\sum_{j=1}^{N/2}\kappa_{2j},
\label{eq:Kodd-Keven}
\end{align}
where the site indices are understood periodically.
The odd bonds and the even bonds each form a complete matching:
within either component, distinct bonds have no common endpoint
and every site belongs to exactly one bond.
The corresponding chirality operators therefore act on
disjoint pairs of sites and mutually commute within each
component.
This decomposition is illustrated in
Fig.~\ref{fig:chirality-bound-structure}a.

This decomposition is useful because the original
nearest-neighbour generator contains overlapping bond terms,
whereas each matching component contains only non-overlapping
terms.
Although commuting matching bonds can still be correlated
through blocks in the chosen \(k\)-producible partition, their
variance can be classified entirely through the block geometries
shown in Fig.~\ref{fig:chirality-bound-structure}.

\vspace{1.0ex}
\subsection*{Step 3: Variance expansion and local bounds}

Consider one of the two matching components,
\(K_{\rm odd}\) or \(K_{\rm even}\), and denote it generically by
\begin{equation}
K_{\mathcal M}
=
\sum_{e\in\mathcal M}\kappa_e ,
\label{eq:KM-def}
\end{equation}
where \(\mathcal M\) denotes the corresponding complete matching.

For Hermitian operators \(A\) and \(B\), we use the symmetrized
covariance
\begin{equation}
\operatorname{Cov}_{\Psi^{(k)}}(A,B)
=
\frac{1}{2}
\left\langle
\Delta A\,\Delta B+\Delta B\,\Delta A
\right\rangle_{\Psi^{(k)}} .
\label{eq:cov-def-main}
\end{equation}
The fluctuation operator is defined by
\begin{equation}
\Delta A
=
A-\langle A\rangle_{\Psi^{(k)}} .
\label{eq:deltaA-def-main}
\end{equation}

The variance of the matching component is then
\begin{equation}
\operatorname{Var}_{\Psi^{(k)}}(K_{\mathcal M})
=
\sum_{e\in\mathcal M}
\operatorname{Var}_{\Psi^{(k)}}(\kappa_e)
+
2
\sum_{\substack{e<f\\e,f\in\mathcal M}}
\operatorname{Cov}_{\Psi^{(k)}}(\kappa_e,\kappa_f).
\label{eq:KM-var-expansion}
\end{equation}

Although
\(\operatorname{Var}_{\Psi^{(k)}}(K_{\mathcal M})\ge0\),
the individual covariance terms in
Eq.~\eqref{eq:KM-var-expansion} can have either sign.
For an upper bound, we therefore use
\begin{equation}
\operatorname{Var}_{\Psi^{(k)}}(K_{\mathcal M})
\le
\sum_{e\in\mathcal M}
\operatorname{Var}_{\Psi^{(k)}}(\kappa_e)
+
2
\sum_{\substack{e<f\\e,f\in\mathcal M}}
\left|
\operatorname{Cov}_{\Psi^{(k)}}(\kappa_e,\kappa_f)
\right|.
\label{eq:KM-var-absolute}
\end{equation}

If two matching bonds touch disjoint sets of blocks in the
chosen \(k\)-producible partition, their covariance vanishes
because the state factorizes between those block sets.

For a single spin-\(1/2\) chirality bond,
\(\operatorname{spec}(\kappa_e)=
\{1/2,-1/2,0,0\}\), and hence
\begin{equation}
\operatorname{Var}_{\Psi^{(k)}}(\kappa_e)
\le\frac{1}{4}.
\label{eq:single-bond-var-main}
\end{equation}
The corresponding internal- and cross-bond geometries are shown
in Fig.~\ref{fig:chirality-bound-structure}c.
For two distinct matching bonds \(e\neq f\), the nonzero
covariances fall into the four block geometries shown in
Fig.~\ref{fig:chirality-bound-structure}d.
The superscripts \({\rm in}\) and \({\rm cr}\) used in the figure
indicate whether a bond is internal or cross with respect to the
reference block \(u\), while \(e\) and \(f\) distinguish the two
matching bonds.
For the internal--internal, internal--cross,
cross--cross with one shared block, and cross--cross with the
same two shared blocks geometries, respectively,
the absolute covariance contributions appearing in
Eq.~\eqref{eq:KM-var-absolute} obey
\begin{equation}
2\left|
\operatorname{Cov}_{\Psi^{(k)}}(\kappa_e,\kappa_f)
\right|
\le
\frac{1}{2},
\quad
\frac{1}{4},
\quad
\frac{1}{8},
\quad
\frac{1}{4}.
\label{eq:local-cov-bounds-main}
\end{equation}
The detailed derivations of the single-bond variance bound and
all four covariance bounds are given in
Supplementary Note {\color{black}S1}.

\vspace{1.0ex}
\subsection*{Step 4: Block-wise allocation}

We now assemble these local estimates block by block.
For each \(u\in\Gamma\), let \(r_u\) be the number of matching
bonds fully contained in \(u\), and let \(c_u\) be the number of
matching bonds with exactly one endpoint in \(u\).
Since \(n_u\) is the number of sites in block \(u\) and
\(\mathcal M\) is a complete matching, every site in \(u\) is the
endpoint of exactly one matching bond.
Consequently,
\begin{equation}
n_u
=
2r_u+c_u,
\qquad
0\le n_u\le k .
\label{eq:nu-def}
\end{equation}
Figure~\ref{fig:chirality-bound-structure}b illustrates these
block parameters for a representative \(N=18\), \(k=9\)
partition.
The partition shown there is only an illustrative example;
the construction below applies to an arbitrary
\(\Gamma\in\mathcal P_k\).

The contribution-to-\(A_u\) columns in
Figs.~\ref{fig:chirality-bound-structure}c,d summarize the
block-wise allocation of the local bounds.
An internal-bond variance is assigned entirely to the block
containing the bond, whereas a cross-bond variance is divided
equally between its two endpoint blocks.
For bond pairs, an internal--internal or internal--cross
contribution associated with \(u\) is assigned to \(u\).
For two cross bonds incident on \(u\), the contribution assigned
to \(u\) is \(1/8\) per unordered pair, irrespective of whether
the two bonds share only \(u\) or share the same two blocks.
In the latter case, the same pair also contributes \(1/8\) to
the other shared block, so that the two allocations together
cover the full \(1/4\) covariance bound.

With this allocation, all variance and covariance contributions
associated with block \(u\) are bounded by
\begin{align}
A_u
&=
\frac{r_u}{4}
+
\frac{c_u}{8}
+
\frac{1}{2}\binom{r_u}{2}
+
\frac{r_uc_u}{4}
+
\frac{1}{8}\binom{c_u}{2}
\notag\\
&=
\frac{r_u^2}{4}
+
\frac{r_uc_u}{4}
+
\frac{c_u(c_u+1)}{16}
\notag\\
&=
\frac{n_u^2+c_u}{16}
\le
\frac{n_u^2+n_u}{16}.
\label{eq:block-allowance}
\end{align}
The first two terms in the first line arise from the
single-bond contributions shown in
Fig.~\ref{fig:chirality-bound-structure}c.
The remaining three terms arise from the bond-pair
contributions shown in
Fig.~\ref{fig:chirality-bound-structure}d.
Here,
\(\binom{r_u}{2}=r_u(r_u-1)/2\) and
\(\binom{c_u}{2}=c_u(c_u-1)/2\)
denote the numbers of unordered pairs of internal bonds within block
\(u\) and of cross bonds with one endpoint in block \(u\), respectively.
In particular, the \(\binom{c_u}{2}\) cross-bond pairs are partitioned
into the two cross--cross geometries shown in
Fig.~\ref{fig:chirality-bound-structure}d.

\vspace{1.0ex}
\subsection*{Step 5: Finite-size maximization}

Equation~\eqref{eq:KM-var-absolute} holds for the arbitrary pure
\(k\)-producible state \(|\Psi^{(k)}\rangle\) and admissible
partition \(\Gamma\in\mathcal P_k\) fixed in Step~1.
We now maximize this state-wise bound over all pure
\(k\)-producible states and all admissible partitions.

To make this maximization explicit, define
\begin{align}
\mathcal V_k(K_{\mathcal M})
&\equiv
\max_{\substack{
\Gamma\in\mathcal P_k\\
|\Psi^{(k)}\rangle
=
\bigotimes_{u\in\Gamma}|\psi_u\rangle
}}
\operatorname{Var}_{\Psi^{(k)}}(K_{\mathcal M}).
\label{eq:Vk-def}
\end{align}
Using Eq.~\eqref{eq:KM-var-absolute} and the block-wise
allocation above, we obtain
\begin{align}
\mathcal V_k(K_{\mathcal M})
&\le
\max_{\substack{
\Gamma\in\mathcal P_k\\
|\Psi^{(k)}\rangle
=
\bigotimes_{u\in\Gamma}|\psi_u\rangle
}}
\Bigg[
\sum_{e\in\mathcal M}
\operatorname{Var}_{\Psi^{(k)}}(\kappa_e)
\notag\\
&\hspace{18mm}
+
2
\sum_{\substack{e<f\\e,f\in\mathcal M}}
\left|
\operatorname{Cov}_{\Psi^{(k)}}(\kappa_e,\kappa_f)
\right|
\Bigg]
\notag\\
&\le
\max_{\Gamma\in\mathcal P_k}
\sum_{u\in\Gamma}A_u
\notag\\
&\le
\frac{1}{16}
\max_{\Gamma\in\mathcal P_k}
\left(
\sum_{u\in\Gamma}n_u^2
+
\sum_{u\in\Gamma}n_u
\right).
\label{eq:matching-var-before-max}
\end{align}

Since the blocks form a partition of all \(N\) sites,
\begin{equation}
\sum_{u\in\Gamma} n_u=N
\qquad
(\Gamma\in\mathcal P_k).
\label{eq:sum-nu-N}
\end{equation}
Under the constraints \(0\le n_u\le k\) and
\(\sum_u n_u=N\), the convexity of \(n_u^2\) implies
\begin{equation}
\max_{\Gamma\in\mathcal P_k}
\sum_{u\in\Gamma} n_u^2
\le
qk^2+s^2.
\label{eq:sum-nu-square}
\end{equation}
The maximal value is obtained by taking as many blocks of size
\(k\) as possible, together with one remaining block of size
\(s\)~\cite{Hyllus2012}.

Equations~\eqref{eq:matching-var-before-max}--%
\eqref{eq:sum-nu-square} give
\begin{equation}
\mathcal V_k(K_{\mathcal M})
\le
\frac{qk^2+s^2+N}{16}.
\label{eq:matching-var-bound}
\end{equation}
Using \(F_Q=4\operatorname{Var}\) for pure states and the
convexity of the QFI for mixed states, we obtain
\begin{equation}
F_Q[\rho^{(k)},K_{\mathcal M}]
\le
\frac{qk^2+s^2+N}{4}.
\label{eq:matching-QFI-bound}
\end{equation}
This result applies separately to both
\(K_{\rm odd}\) and \(K_{\rm even}\).

\vspace{1.0ex}
\subsection*{Step 6: Recombination of the two matchings}

Finally, the square root of the QFI obeys the triangle inequality
with respect to the generator,
\begin{align}
\sqrt{F_Q[\rho^{(k)},K_{\rm tot}]}
&\le
\sqrt{F_Q[\rho^{(k)},K_{\rm odd}]}
\notag\\
&\quad+
\sqrt{F_Q[\rho^{(k)},K_{\rm even}]}.
\label{eq:QFI-triangle-main}
\end{align}
A proof of Eq.~\eqref{eq:QFI-triangle-main} is given in
Supplementary Note {\color{black}S2}.
Substituting Eq.~\eqref{eq:matching-QFI-bound} for both
components and squaring yields
\begin{equation}
F_Q[\rho^{(k)},K_{\rm tot}]
\le
qk^2+s^2+N,
\label{eq:Ktot-QFI-finite}
\end{equation}
which is equivalent to
Eq.~\eqref{eq:kplusone-bound-finite}.

For a calculated or experimentally extracted QFI density
\(f_Q=F_Q[\rho,K_{\rm tot}]/N\), the largest integer \(k\)
for which
\begin{equation}
f_Q
>
k+1-\frac{s(k-s)}{N},
\qquad
N=qk+s,
\label{eq:depth-test}
\end{equation}
is satisfied provides the certification
\(\text{entanglement depth}\ge k+1\).

The bound in Eq.~\eqref{eq:kplusone-bound-finite} is rigorous
but should not be interpreted as tight.
In particular, Eq.~\eqref{eq:QFI-triangle-main} replaces the
joint optimization of \(K_{\rm odd}+K_{\rm even}\) by two
independent matching bounds and therefore discards correlations
and simultaneous-saturation constraints between the odd and
even sectors.
A joint optimization retaining this information could lead to
a tighter \(k\)-producible threshold.



\section{Multipartite entanglement in an extended \(J_1\)--\(J_2\) chain}

\begin{figure*}[t]
\centering
\includegraphics[
width=\textwidth
]{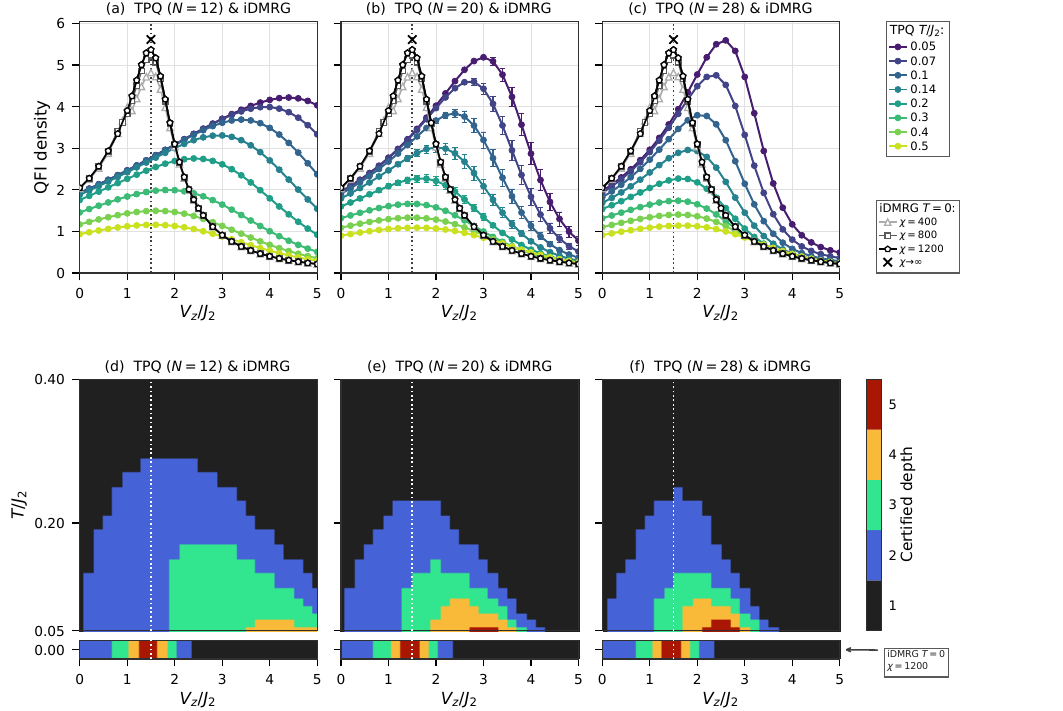}

\caption{
Zero and finite-temperature chirality-QFI densities
\(f_Q^{(0)}(R)\) and \(f_Q^{(T)}(N)\), defined in
Eqs.~\eqref{eq:fq-zero-R} and \eqref{eq:fq-finite-T},
respectively, obtained from iDMRG and TPQ calculations.
\textbf{a--c}, \(V_z/J_2\) dependence of the finite-temperature
chirality-QFI density \(f_Q^{(T)}(N)\) at representative fixed
temperatures for \(N=12\), \(20\), and \(28\), respectively,
together with the \(T=0\) iDMRG results.
Error bars, where available, denote the jackknife standard error
of the ratio-of-sums TPQ estimate.
The iDMRG results are evaluated with \(R=400\) and bond
dimensions \(\chi=400\), \(800\), and \(1200\).
The cross at \(V_z/J_2=1.5\) denotes the extrapolated
\(\chi\to\infty\) QFI density, \(f_Q^{(0)}\sim 5.61\), obtained from
the \(1/\chi\) dependence up to
\(\chi=2400\).
\textbf{d--f}, Certified entanglement depth obtained from
\(f_Q^{(T)}(N)\) in the \(V_z/J_2\)--\(T/J_2\) plane for
\(N=12\), \(20\), and \(28\), respectively, using the
entanglement-depth criterion in
Eq.~\eqref{eq:kplusone-bound-simple}.
The narrow strips below the maps show the certified
entanglement depth inferred from the \(T=0\) iDMRG QFI density
for \(\chi=1200\).
The colour scale directly represents the certified
entanglement depth, with the colours labelled \(1\) through
\(5\) corresponding to the respective certified depths.
The vertical dotted line marks \(V_z/J_2=1.5\).
With increasing system size, the centre of the
finite-temperature QFI-enhanced region shifts towards the
iDMRG maximum near \(V_z/J_2\simeq1.5\).
}

\label{fig:qfi-tpq-idmrg}
\end{figure*}

We now apply the chirality-QFI criterion in
Eq.~\eqref{eq:kplusone-bound-simple} to the extended spin-\(1/2\)
\(J_1\)--\(J_2\) XXZ chain introduced by Furukawa {\it et al}. for
one-dimensional spin-\(1/2\) multiferroics~\cite{Furukawa2008}.
The model is defined by
\begin{equation}
H = H_{J_1J_2}+H_{\rm BDM},
\label{eq:extended-H}
\end{equation}
\begin{equation}
H_{J_1J_2}
=
\sum_{n=1}^{2}\sum_j
J_n
\left[
\mathbf{S}_j\cdot\mathbf{S}_{j+n}
+
(\Delta-1)S_j^zS_{j+n}^z
\right],
\label{eq:J1J2-H}
\end{equation}
and
\begin{equation}
H_{\rm BDM}
=
-V_z\sum_j\kappa_j\kappa_{j+2},
\qquad
\kappa_j=(\mathbf{S}_j\times\mathbf{S}_{j+1})^z.
\label{eq:BDM-H}
\end{equation}
The last term is the biquadratic Dzyaloshinskii--Moriya interaction
generated by integrating out transverse optical phonons coupled to the
local vector chirality~\cite{Onoda2007,Furukawa2008}.
We use the same parameter set as Furukawa {\it et al}.,
\begin{equation}
\frac{J_1}{J_2}=-0.1,
\qquad
\Delta=0.9,
\qquad
0\leq\frac{V_z}{J_2}\leq5.
\label{eq:model-parameters}
\end{equation}

Along this parameter cut, Furukawa {\it et al}. found, using exact
diagonalization and finite-size scaling based on the \(\epsilon\)
algorithm, that vector-chiral order is already present at
\(V_z/J_2=0\), although it is weak there, and becomes stronger as
\(V_z/J_2\) is increased~\cite{Furukawa2008}.
We therefore regard the evolution between the weakly ordered and
strongly ordered vector-chiral regimes as a crossover, occurring around
\(V_z/J_2\simeq1.5\).

At zero temperature, the QFI for \(T=0\) reported below is evaluated
using iDMRG implemented with the TeNPy library~\cite{TeNPy01}.
In the vector-chiral phase, the two opposite chiral sectors are related
by the broken \(\mathbb{Z}_2\) symmetry.
To select a single chiral sector and avoid artificial switching or
cat-state contributions in the infinite-MPS calculation, we add a weak
uniform chiral field to Eq.~\eqref{eq:extended-H},
\begin{equation}
H_\kappa=-h_\kappa K_{\rm tot},
\qquad
\frac{h_\kappa}{J_2}=10^{-4}.
\label{eq:chiral-field}
\end{equation}
With this sign convention, \(h_\kappa>0\) selects the positive-chirality
sector.

For a pure ground state, the QFI is four times the variance of the
generator.
For an \(N\)-site system, we define the ground-state chirality-QFI
density as
\begin{equation}
f_Q^{(0)}(N)
\equiv
\frac{F_Q[|\Psi_0\rangle,K_{\rm tot}]}{N}
=
\frac{4}{N}
\left(
\langle K_{\rm tot}^2\rangle
-
\langle K_{\rm tot}\rangle^2
\right).
\label{eq:fq-zero-N}
\end{equation}
The ground-state QFI density in the thermodynamic limit is then
\begin{equation}
f_Q^{(0)}
\equiv
\lim_{N\to\infty}f_Q^{(0)}(N).
\label{eq:fq-zero-thermo}
\end{equation}

For the iDMRG evaluation, we use the equal-time connected chirality
correlations of the resulting pure, sector-selected infinite
matrix-product state.
We choose a four-site unit cell, \(L_{\rm cell}=4\), which is
appropriate for capturing the vector-chiral state considered here.
Let the bond index \(a=0,\ldots,L_{\rm cell}-1\) label the
vector-chirality bonds inside a unit cell, and let \(\kappa_{a+r}\)
denote the bond obtained by shifting \(\kappa_a\) by \(r\) bond
spacings along the infinite chain.
We define the unit-cell averaged connected chirality correlator
\begin{equation}
C_{\rm c}(r)
=
\frac{1}{L_{\rm cell}}
\sum_{a=0}^{L_{\rm cell}-1}
{\rm Re}
\left[
\langle\kappa_a\kappa_{a+r}\rangle
-
\langle\kappa_a\rangle
\langle\kappa_{a+r}\rangle
\right].
\label{eq:Cc-idmrg}
\end{equation}
The real part gives the symmetrized connected covariance entering the
variance of the Hermitian operator \(K_{\rm tot}\).
In practice, we compute the accumulated quantity
\begin{equation}
f_Q^{(0)}(R)
:=
4
\left[
C_{\rm c}(0)
+
2\sum_{r=1}^{R}C_{\rm c}(r)
\right].
\label{eq:fq-zero-R}
\end{equation}
The ground-state QFI density in the thermodynamic limit is obtained from the
large-\(R\) saturation of this accumulated quantity:
\begin{equation}
f_Q^{(0)}
=
\lim_{R\to\infty}f_Q^{(0)}(R)
=
\lim_{N\to\infty}f_Q^{(0)}(N).
\label{eq:fq-zero-limits}
\end{equation}
Unless otherwise stated, we use \(R=400\), and convergence was checked
up to \(R=600\).

At finite temperature, we evaluate the chirality-QFI density using
thermal pure quantum states.
For periodic chains with \(N=12\), \(20\), and \(28\) sites, we compute
the real-time response associated with \(K_{\rm tot}\).
For the thermal state
\begin{equation}
\rho(T)
=
\frac{e^{-\beta H}}{{\rm Tr}\,e^{-\beta H}},
\qquad
\beta=\frac{1}{T},
\label{eq:thermal-rho}
\end{equation}
the QFI density associated with \(K_{\rm tot}\) can be expressed in
terms of the dissipative part of the dynamical susceptibility as
\begin{align}
f_Q^{(T)}(N)
&\equiv
\frac{F_Q[\rho(T),K_{\rm tot}]}{N}
\notag\\
&=
\frac{4}{\pi N}
\int_0^\infty
d\omega\,
\tanh\left(\frac{\beta\omega}{2}\right)
\chi_{KK}''(\omega,T).
\label{eq:fq-finite-T}
\end{align}
Here \(\chi_{KK}''(\omega,T)\) denotes the imaginary part of the causal
susceptibility for \(K_{\rm tot}\).
We define
\begin{equation}
\chi_{KK}(t)
=
i\theta(t)
\left\langle
[K_{\rm tot}(t),K_{\rm tot}(0)]
\right\rangle_T,
\label{eq:chiKK-time}
\end{equation}
where \(\theta(t)\) is the Heaviside step function,
\(K_{\rm tot}(t)=e^{iHt}K_{\rm tot}e^{-iHt}\), and
\(\langle O\rangle_T={\rm Tr}[\rho(T)O]\).
Its Fourier transform is
\begin{align}
\chi_{KK}(\omega,T)
&=
\int_0^\infty
dt\,
e^{i\omega t}
\chi_{KK}(t),
\label{eq:chiKK-frequency}
\\
\chi_{KK}''(\omega,T)
&=
{\rm Im}\,\chi_{KK}(\omega,T).
\label{eq:chiKK-imaginary-part}
\end{align}

In practice, we compute the dynamical response associated with
\(K_{\rm tot}\) in the TPQ framework and insert the resulting
\(\chi_{KK}''(\omega,T)\) into Eq.~\eqref{eq:fq-finite-T};
details of the TPQ construction, real-time evolution, Fourier
transform, and TPQ-sample averaging are provided in the
Supplementary Information.

The numerical results are summarized in Fig.~\ref{fig:qfi-tpq-idmrg}.
The upper panels, Figs.~\ref{fig:qfi-tpq-idmrg}a--c, show the
\(V_z/J_2\) dependence of the chirality-QFI density obtained from
the finite-temperature TPQ calculations together with the \(T=0\)
iDMRG results.
The three panels differ only in the system size used in the TPQ
calculations, \(N=12\), \(20\), and \(28\), respectively, whereas
the same iDMRG ground-state data are shown in all three panels.

The previous study by Furukawa {\it et al}.~\cite{Furukawa2008} identified
a crossover around \(V_z/J_2\simeq1.5\) from the weakly ordered
vector-chiral regime to the strongly ordered vector-chiral regime.
The iDMRG results show that the ground-state chirality-QFI density
develops a pronounced peak around the crossover at
\(V_z/J_2\simeq1.5\).
The height of the peak around \(V_z/J_2\simeq1.5\) retains a small
bond-dimension dependence, but the results for
\(\chi=400\), \(800\), and \(1200\) show a clear tendency toward
convergence.
At \(V_z/J_2=1.5\), an extrapolation in \(1/\chi\) gives an estimated
QFI density of approximately \(5.61\) in the
\(\chi\rightarrow\infty\) limit.

The TPQ results exhibit a substantial finite-size dependence.
Nevertheless, as the system size is increased from \(N=12\) to
\(20\) and then to \(28\), both the position and the height of the
finite-temperature QFI peak progressively approach those of the
\(T=0\) iDMRG result.
The lower panels, Figs.~\ref{fig:qfi-tpq-idmrg}d--f, translate the
calculated QFI densities into certified entanglement depths using the
chirality-QFI entanglement criterion of
Eq.~\eqref{eq:kplusone-bound-simple}.
The color maps show the certified entanglement depth
obtained from the finite-size, finite-temperature TPQ calculations
for \(N=12\), \(20\), and \(28\), while the narrow strips below
them show the corresponding \(T=0\) result obtained from iDMRG.
The same iDMRG data are used for all three strips.
As the system size increases, regions with certified entanglement
depth \(5\) emerge, and the region with larger certified depth
progressively shifts toward \(V_z/J_2\simeq1.5\).

Together with the \(T=0\) iDMRG result, which directly establishes
multipartite entanglement in the thermodynamic limit, these
finite-size TPQ trends strongly suggest that multipartite
entanglement encoded in the vector-chiral correlations around
\(V_z/J_2\simeq1.5\) also survives at finite temperatures in the
thermodynamic limit.

\section{Experimental access via dielectric spectroscopy}
\label{sec:exp}

We finally discuss how the dynamical vector-chirality response entering
the QFI can, in principle, be accessed experimentally through dielectric
spectroscopy. Within the spin-current, or inverse
Dzyaloshinskii--Moriya, mechanism, the electric dipole associated with a
bond \(ij\) is written as~\cite{Katsura2005,Katsura2007,Tokura2010}
\begin{equation}
\mathbf{d}_{ij}
=
g_{\rm sc}\,
\mathbf{e}_{ij}
\times
\left(
\mathbf{S}_i\times\mathbf{S}_j
\right),
\label{eq:spin-current-dipole}
\end{equation}
where \(\mathbf{e}_{ij}\) is the unit vector along the bond and
\(g_{\rm sc}\) denotes the magnetoelectric coupling coefficient.
Taking the chain direction to be the \(x\) axis,
\(\mathbf{e}_{i,i+1}=\hat{\mathbf{x}}\), the \(z\) component of the
vector chirality,
\begin{equation}
\kappa_i^z
=
\left(
\mathbf{S}_i\times\mathbf{S}_{i+1}
\right)^z,
\end{equation}
generates an electric dipole along the \(y\) direction. Absorbing the
overall sign associated with
\(\hat{\mathbf{x}}\times\hat{\mathbf{z}}=-\hat{\mathbf{y}}\)
into the coupling constant, we write
\begin{equation}
d_{i,i+1}^{y}
=
\lambda\,\kappa_i^z.
\label{eq:bond-dipole-chirality}
\end{equation}
If the same coupling \(\lambda\) applies uniformly to all bonds, the
total electric dipole along the \(y\) direction becomes
\begin{equation}
D_y
=
\sum_i d_{i,i+1}^{y}
=
\lambda
\sum_i\kappa_i^z
=
\lambda K_{\rm tot}.
\label{eq:total-dipole-chirality}
\end{equation}

Equation~\eqref{eq:total-dipole-chirality} directly relates the
corresponding dynamical response functions. In particular,
\begin{equation}
\chi_{D_yD_y}''(\omega,T)
=
\lambda^2
\chi_{KK}''(\omega,T),
\label{eq:dipole-chirality-response}
\end{equation}
where \(\omega\) denotes the energy transfer, consistently with the
convention \(\hbar=1\) used in the theoretical and TPQ calculations.
The quantity \(\chi_{KK}''(\omega,T)\) is the same dynamical
vector-chirality susceptibility defined in
Eq.~\eqref{eq:chiKK-imaginary-part}.
A direct relation between the dielectric response and the dynamical
vector-chirality correlation has also been discussed for frustrated
spin chains~\cite{Furukawa2008,Furukawa2010}.

To connect Eq.~\eqref{eq:dipole-chirality-response} to a dielectric
measurement, we denote by \(\varepsilon_{yy}(\omega,T)\) the complex
relative permittivity for an electric field polarized along the
\(y\) direction, and by
\(\varepsilon_{yy,\chi}''(\omega,T)\) the contribution to its imaginary
part, i.e., to the total dielectric loss
\(\varepsilon_{yy}''(\omega,T)\), arising from the vector-chirality
electric-dipole channel. For this isolated channel,
Eq.~\eqref{eq:dipole-chirality-response} gives
\begin{equation}
\varepsilon_{yy,\chi}''(\omega,T)
=
A\,
\frac{\chi_{KK}''(\omega,T)}{N},
\qquad
A
=
\frac{\lambda^2}{\varepsilon_0 v_s},
\label{eq:dielectric-chirality-response}
\end{equation}
where \(\varepsilon_0\) is the vacuum permittivity and
\(v_s=V/N\) is the volume per magnetic spin, with \(V\) the sample
volume and \(N\) the number of magnetic spins.

Thus, apart from the material-dependent overall conversion factor
\(A\), the vector-chirality contribution to the dielectric-loss
spectrum directly probes the same quantity
\(\chi_{KK}''(\omega,T)/N\) that is evaluated in our TPQ calculations.
Since this susceptibility determines the finite-temperature chirality
QFI through Eq.~\eqref{eq:fq-finite-T}, once
\(\varepsilon_{yy,\chi}''(\omega,T)\) is experimentally isolated and
the conversion factor \(A\) is determined, the chirality-QFI density
can in principle be reconstructed from dielectric spectroscopy as
\begin{equation}
f_Q^{(T)}
=
\frac{4}{\pi A}
\int_0^\infty
d\omega\,
\tanh\left(
\frac{\omega}{2T}
\right)
\varepsilon_{yy,\chi}''(\omega,T).
\label{eq:fq-from-dielectric-loss}
\end{equation}

Dielectric spectroscopy has already been used to resolve chiral
dynamics in the quasi-one-dimensional frustrated spin-chain compound
LiCuVO$_4$, where background-corrected dielectric-loss spectra were
associated with chiral fluctuations and low-energy chiral
excitations~\cite{Grams2022}. In practice, the material-dependent issues
are therefore the identification of the vector-chirality contribution
\(\varepsilon_{yy,\chi}''(\omega,T)\) within the total dielectric loss
\(\varepsilon_{yy}''(\omega,T)\) and the determination of the overall
factor \(A\).

A possible route to determine \(A\) independently is to calibrate the
magnetoelectric coupling \(\lambda\) using the static electric
polarization. From Eq.~\eqref{eq:total-dipole-chirality}, the
polarization density satisfies
\begin{equation}
P_y
=
\frac{\langle D_y\rangle}{V}
=
\frac{\lambda}{v_s}
\frac{\langle K_{\rm tot}\rangle}{N}.
\label{eq:lambda-calibration}
\end{equation}
The electric polarization can be measured quantitatively by
pyrocurrent or magnetocurrent measurements. In LiCuVO$_4$, for example,
quantitative measurements of the ferroelectric polarization and its
temperature- and magnetic-field dependence have been reported
~\cite{Schrettle2008,Ruff2019,Grams2019}.
A low-temperature saturation polarization of approximately
\(P_s\simeq30\,\mu{\rm C/m^2}\) has been reported in the
three-dimensionally ordered spin-spiral phase
~\cite{Ruff2019,Grams2019}, demonstrating that the absolute
polarization scale can be determined experimentally.
Importantly, Ruff \textit{et al.} also observed a
smaller but finite polarization of order \(0.1P_s\) above the
three-dimensional ordering temperature \(T_N\simeq2.5\,{\rm K}\), in a
field-induced vector-chiral regime without long-range spin
order~\cite{Ruff2019}. Related chiral fluctuations above \(T_N\) have
also been resolved dynamically by dielectric
spectroscopy~\cite{Grams2022}.

A practical calibration procedure could therefore proceed as follows.
(i) An effective one-dimensional spin Hamiltonian appropriate to the
material is first constrained independently by comparing calculated
thermodynamic and magnetic observables, such as the specific heat,
magnetic susceptibility, magnetization, and, where available,
spin-correlation functions, with experiment.
(ii) For the resulting effective Hamiltonian, the vector chirality
\(\langle K_{\rm tot}\rangle/N\) is evaluated by DMRG in the
sector-selected ground state, including the same external magnetic
field and relevant anisotropies as in the polarization experiment.
(iii) Experimentally, a definite chiral sector is first selected, for
example by electric-field cooling, and the sample is then brought to
a temperature slightly above \(T_N\), where the three-dimensional spin
order has disappeared but the vector-chiral polarization remains
finite. This regime is a natural candidate for comparison with an
effective one-dimensional spin Hamiltonian, provided that interchain
effects are sufficiently weak in this temperature range.
(iv) The polarization \(P_y\) is then measured quantitatively in this
low-temperature vector-chiral regime.
(v) Provided that thermal corrections to the intrachain chirality are
sufficiently small, comparison with the DMRG value gives
\begin{equation}
\lambda
\simeq
v_s
\frac{P_y^{\rm exp}}
{\langle K_{\rm tot}\rangle_{\rm DMRG}/N},
\label{eq:lambda-calibration-DMRG}
\end{equation}
where \(P_y^{\rm exp}\) is evaluated at a temperature slightly above
\(T_N\), in the vector-chiral regime where the three-dimensional spin
order has disappeared but the finite polarization remains observable.
This then determines
\(A=\lambda^2/(\varepsilon_0v_s)\).

Measuring the ratio in Eq.~\eqref{eq:lambda-calibration-DMRG} for several magnetic fields might provide an additional consistency check: an approximately
field-independent value of \(\lambda\) would support both the
effective one-dimensional spin model and the assumed linear relation
\(D_y=\lambda K_{\rm tot}\). Once \(A\) is calibrated and the
vector-chirality contribution to the dielectric loss is isolated,
Eq.~\eqref{eq:fq-from-dielectric-loss} provides an experimental route
to multipartite-entanglement certification from vector-chiral
correlations in quasi-one-dimensional multiferroic quantum magnets.

\section{Discussion}

The present results suggest a broader way of using quantum Fisher
information in quantum many-body systems.  Recent approaches based on
reduced density matrices, genuine multipartite entanglement, and
semidefinite programming have made it possible to characterize the
depth and spatial structure of entanglement in strongly correlated
states, including quantum spin liquids~\cite{Lyu2026,Sid2025v2}.
Such approaches provide detailed information about how entanglement is
distributed among subsets of degrees of freedom.  A complementary
question, particularly natural from the viewpoint of condensed-matter
physics, is which physically meaningful correlation channel carries
that entanglement.  Because the QFI depends on the choice of generator,
a generator constructed from a familiar local or composite observable
provides a way of addressing this question directly.  In the present
case, choosing the total vector chirality as the generator makes the
resulting witness sensitive to multipartite entanglement encoded in
vector-chiral correlations, rather than only establishing the presence
of entanglement as an abstract property of the many-body state.

An important feature of the bound derived here is that it does not
depend on the Hamiltonian used in Sec.~III.  For an even spin-$1/2$
periodic chain, the bound follows from the local algebra and overlap
structure of the nearest-neighbour vector-chirality operators entering
$K_{\rm tot}$.  It can therefore be applied without modification to
other spin-$1/2$ chain Hamiltonians whenever the same generator is
considered.  Extending the construction to other spin magnitudes,
lattice geometries, or composite generators will in general require
new $k$-producible bounds. 
As shown in Sec.~II, the present derivation follows a systematic
sequence that may provide a route to more general composite
observables.
One first reduces the problem to pure block-product states using
the convexity of the QFI (Step 1), decomposes the overlap structure
of the composite generator into tractable subsets (Step 2), and
characterizes the local operator algebra to bound the corresponding
variances and covariances (Step 3).
These local bounds are then organized into block-wise contributions
(Step 4) and maximized under the $k$-producibility constraint
(Step 5), before the separately bounded components are recombined
(Step 6).
In this sense, the strategy used here should be adaptable in principle
to a substantially broader class of observables and lattices.

A limitation of this strategy, however, is that obtaining a tight
$k$-producible bound may be difficult, because decomposing the
generator into tractable subsets and bounding them separately can
discard constraints that act jointly across those subsets.
In the present case, this loss is particularly explicit in Step 6:
the recombination of the odd and even matchings through the QFI
triangle inequality discards simultaneous constraints between the
two sectors.
Accordingly, the resulting $k+1$ bound is rigorous but is not
expected to be tight, leaving room for tighter bounds based on a
joint optimization.

This freedom in the choice of generator opens a number of
possible directions.
Dimer or bond observables are natural in frustrated quantum
magnets with valence-bond physics.
Potential settings include the spontaneously dimerized regime
of the antiferromagnetic spin-$1/2$ $J_1$--$J_2$ chain beyond
the fluid--dimer transition, as well as dimerized phases in
frustrated ferromagnetic $J_1$--$J_2$ XXZ chains
~\cite{Okamoto1992,Hikihara2001,Furukawa2012}.
Related possibilities arise in extended regimes of the
Shastry--Sutherland model beyond its exact dimer-product phase,
where competing plaquette and magnetically ordered phases appear
~\cite{Koga2000}.
Composite bond correlations are also particularly intriguing
in the Kitaev honeycomb model, where bond--bond correlations
show algebraic behaviour in the gapless phase despite the
short-ranged nature of conventional spin correlations
~\cite{Yang2008}.
Spin-nematic or quadrupolar observables provide another natural
class, as realized in frustrated ferromagnetic $J_1$--$J_2$
chains and square-lattice frustrated ferromagnets
~\cite{Hikihara2008,Shannon2006}.
Scalar spin chirality is similarly a central observable in
chiral spin-liquid states, with prominent examples provided by
frustrated kagome- and triangular-lattice models
~\cite{He2014,Bauer2014,Hu2016,Wietek2017}.
Their QFI would probe multipartite entanglement encoded in
correlation channels that are different from both conventional
dipolar spin correlations and the vector-chiral correlations
studied here.
More generally, choosing an observable that is physically
characteristic of the state or regime of interest may provide
an ``observable-resolved'' view of multipartite entanglement.
Whenever the dynamical response of that observable is
experimentally accessible, this construction also offers, in
principle, a route for connecting entanglement certification
to the spectroscopy most naturally associated with the
relevant degrees of freedom.

We hope that extending this strategy to
other observables, materials, and experimental probes will stimulate
a broader interplay between quantum-information concepts and the
correlation functions routinely used to characterize quantum
materials.

\section{Summary}

We have established a multipartite-entanglement criterion based on
the QFI of the total nearest-neighbour vector chirality.
For spin-$1/2$ periodic chains with an even number $N\ge4$ of sites,
we derived the rigorous $k$-producible bound $f_Q\le k+1$,
together with its finite-size refinement.
Violating the corresponding bound certifies entanglement depth
of at least $k+1$.
The criterion depends on the structure of the generator rather
than on a particular Hamiltonian.

Applying this criterion to an extended $J_1$--$J_2$ XXZ chain,
we demonstrated multipartite entanglement encoded in vector-chiral
correlations at zero and finite temperatures using iDMRG and TPQ
calculations, respectively.
We also related the dynamical response entering the chirality QFI
to dielectric loss through the spin-current mechanism, providing
a route, in principle, to experimental entanglement certification
in multiferroic spin-chain magnets.
Together, these results extend QFI-based entanglement diagnostics
beyond one-site spin generators and provide a concrete example of
how higher-order spin correlations can be used to characterize
and experimentally access multipartite entanglement in quantum
materials.

\section*{Acknowledgements}

This work was supported by the Theory of Quantum Matter Unit of the Okinawa Institute of Science and Technology Graduate University (OIST), and was supported by JSPS KAKENHI Grant-in-Aid for Scientific Research (C), Grant No.~25K07213, and by MEXT KAKENHI Grant-in-Aid for Transformative Research Areas A ``Extreme Universe,'' Grant Nos.~22H05266 and 24H00974.
The author gratefully acknowledges helpful discussions with Shunsuke Furukawa and Ryuji Takagi, and thanks Snigdh Sabharwal, Matthias Gohlke, and Nic Shannon for carefully reading the manuscript and for their useful comments.
The iDMRG calculations were performed using the TeNPy library~\cite{TeNPy01}.
OpenAI Codex (GPT-5.6 Sol) was used to assist with modifications and debugging of the TeNPy-based iDMRG implementation.
ChatGPT (OpenAI, GPT-5.6 Sol) was used as an interactive aid in developing and checking the mathematical derivation presented in Supplementary Note S1.6, conducting literature searches, and assisting with English-language polishing of the manuscript.
All AI-assisted mathematical derivations, code modifications, numerical procedures, and manuscript text were independently reviewed and verified by the author, who takes full responsibility for the content of this work.
The finite-temperature TPQ calculations were performed using a code developed by the author based on TITPACK Ver.~2~\cite{Nishimori1991TITPACK}.
Numerical calculations were carried out using HPC facilities provided by the Supercomputing Center, ISSP, the University of Tokyo; the Yukawa Institute Computer Facility at Kyoto University (YITP); OIST; and the supercomputer Fugaku, provided by RIKEN through the HPCI System Research Project (Project IDs: hp250047, hp260092 and hp260309).
The author thanks Hiroki Nakano for assistance with the use of Fugaku.

\newpage

\newpage

\section*{Supplementary Note S1:
Detailed derivation of the \(k\)-producible chirality-QFI bound}

This note provides the detailed derivations underlying
Steps 3--5 of Sec.~II.
We consider an even spin-\(1/2\) periodic chain and a fixed
integer \(1\le k\le N\).
For an arbitrary pure \(k\)-producible state, we choose an
admissible block partition
\begin{equation}
|\Psi^{(k)}\rangle
=
\bigotimes_{u\in\Gamma}|\psi_u\rangle,
\qquad
n_u\le k,
\label{eq:S1-kprod-state}
\end{equation}
where \(n_u\) is the number of sites in block \(u\).
The blocks need not be spatially contiguous, and no assumption
is made that the state within a block is genuinely
\(n_u\)-partite entangled.

We consider one complete matching \(\mathcal M\), corresponding
to either the odd or the even matching of the main text, and
write
\begin{equation}
K_{\mathcal M}
=
\sum_{e\in\mathcal M}\kappa_e .
\label{eq:S1-KM}
\end{equation}
For each \(u\in\Gamma\), let \(r_u\) denote the number of
matching bonds whose two endpoints both belong to \(u\), and
let \(c_u\) denote the number of matching bonds with exactly
one endpoint in \(u\).
Because \(\mathcal M\) is a complete matching, every site in
\(u\) is the endpoint of exactly one matching bond, and hence
\begin{equation}
n_u=2r_u+c_u .
\label{eq:S1-coverage}
\end{equation}
Since the blocks form a partition of the \(N\) sites,
\begin{equation}
\sum_{u\in\Gamma}n_u=N.
\label{eq:S1-sum-nu}
\end{equation}

Throughout this note,
\(\operatorname{Var}_{\Psi^{(k)}}\) and
\(\operatorname{Cov}_{\Psi^{(k)}}\) refer to the fixed pure
state in Eq.~\eqref{eq:S1-kprod-state}.
For Hermitian operators \(A\) and \(B\), we use the
symmetrized covariance
\begin{equation}
\operatorname{Cov}_{\Psi^{(k)}}(A,B)
=
\frac12
\left\langle
\Delta A\,\Delta B+\Delta B\,\Delta A
\right\rangle_{\Psi^{(k)}} ,
\label{eq:S1-cov-def}
\end{equation}
with
\begin{equation}
\Delta A
=
A-\langle A\rangle_{\Psi^{(k)}} .
\label{eq:S1-delta-def}
\end{equation}

\subsection*{S1.1 Single-bond chirality algebra and covariance inequality}

For a spin-\(1/2\) bond with the ordered endpoints \((i,j)\),
the vector-chirality operator is
\begin{equation}
\kappa_{i,j}
=
S_i^xS_j^y-S_i^yS_j^x.
\label{eq:S1-kappa-def}
\end{equation}
Using
\(S^\pm=S^x\pm iS^y\), this can be written as
\begin{equation}
\kappa_{i,j}
=
\frac{1}{2i}
\left(
S_i^-S_j^+
-
S_i^+S_j^-
\right).
\label{eq:S1-kappa-ladder}
\end{equation}
In the two-site \(S^z\) basis
\begin{equation}
\left(
|\uparrow\uparrow\rangle,
|\uparrow\downarrow\rangle,
|\downarrow\uparrow\rangle,
|\downarrow\downarrow\rangle
\right),
\label{eq:S1-two-site-basis}
\end{equation}
the matrix representation is
\begin{equation}
\kappa_{i,j}
=
\begin{pmatrix}
0&0&0&0\\
0&0&i/2&0\\
0&-i/2&0&0\\
0&0&0&0
\end{pmatrix}.
\label{eq:S1-kappa-matrix}
\end{equation}
Its spectrum is therefore
\begin{equation}
\operatorname{spec}(\kappa_{i,j})
=
\left\{
-\frac12,0,0,\frac12
\right\}.
\label{eq:S1-kappa-spectrum}
\end{equation}

Squaring Eq.~\eqref{eq:S1-kappa-matrix} gives
\begin{equation}
\kappa_{i,j}^2
=
\begin{pmatrix}
0&0&0&0\\
0&1/4&0&0\\
0&0&1/4&0\\
0&0&0&0
\end{pmatrix},
\label{eq:S1-kappa-square-matrix}
\end{equation}
or equivalently
\begin{equation}
\kappa_{i,j}^2
=
\frac18 I
-
\frac12 S_i^zS_j^z
=
\frac18
\left(
I-\sigma_i^z\sigma_j^z
\right).
\label{eq:S1-kappa-square}
\end{equation}

{\color{black}
Consequently, for an arbitrary two-site state \(\rho\),
the variance can also be written exactly as
\begin{align}
\operatorname{Var}_{\rho}(\kappa_{i,j})
&=
\langle\kappa_{i,j}^2\rangle_{\rho}
-
\langle\kappa_{i,j}\rangle_{\rho}^{2}
\notag\\
&=
\frac18
-
\frac12\langle S_i^zS_j^z\rangle_{\rho}
-
\langle\kappa_{i,j}\rangle_{\rho}^{2}.
\label{eq:S1-kappa-var-exact}
\end{align}

We now derive the state-independent variance bound directly
from the largest and smallest eigenvalues of the operator.
Let \(A\) be an arbitrary Hermitian operator with spectral
decomposition
\begin{equation}
A
=
\sum_n\lambda_n|n\rangle\langle n|.
\label{eq:S1-range-spectral-decomp}
\end{equation}
For an arbitrary density matrix \(\rho\), define
\begin{equation}
p_n
=
\langle n|\rho|n\rangle.
\label{eq:S1-range-pn}
\end{equation}
Then
\begin{equation}
p_n\ge0,
\qquad
\sum_n p_n=1,
\end{equation}
and
\begin{equation}
\mu
:=
\langle A\rangle_{\rho}
=
\sum_n p_n\lambda_n.
\label{eq:S1-range-mean}
\end{equation}
The variance is
\begin{align}
\operatorname{Var}_{\rho}(A)
&=
\sum_n p_n\lambda_n^2-\mu^2
\notag\\
&=
\sum_n p_n(\lambda_n-\mu)^2.
\label{eq:S1-range-variance}
\end{align}

Let \(\lambda_{\min}\) and \(\lambda_{\max}\) denote the
minimum and maximum eigenvalues of \(A\).
Since
\begin{equation}
\lambda_{\min}
\le
\lambda_n
\le
\lambda_{\max},
\end{equation}
we have
\begin{equation}
(\lambda_n-\lambda_{\min})
(\lambda_{\max}-\lambda_n)
\ge0.
\end{equation}
Equivalently,
\begin{equation}
\lambda_n^2
\le
(\lambda_{\min}+\lambda_{\max})\lambda_n
-
\lambda_{\min}\lambda_{\max}.
\end{equation}
Multiplying by \(p_n\) and summing over \(n\) yields
\begin{equation}
\sum_n p_n\lambda_n^2
\le
(\lambda_{\min}+\lambda_{\max})\mu
-
\lambda_{\min}\lambda_{\max}.
\label{eq:S1-range-second-moment}
\end{equation}
Therefore
\begin{align}
\operatorname{Var}_{\rho}(A)
&=
\sum_n p_n\lambda_n^2-\mu^2
\notag\\
&\le
(\lambda_{\max}-\mu)
(\mu-\lambda_{\min}).
\label{eq:S1-range-product}
\end{align}
Because
\begin{equation}
\lambda_{\min}\le\mu\le\lambda_{\max},
\end{equation}
the right-hand side of
Eq.~\eqref{eq:S1-range-product} is maximized at
\begin{equation}
\mu
=
\frac{\lambda_{\min}+\lambda_{\max}}{2},
\end{equation}
which gives
\begin{equation}
0
\le
\operatorname{Var}_{\rho}(A)
\le
\frac{
\left(
\lambda_{\max}-\lambda_{\min}
\right)^2
}{4}.
\label{eq:S1-spectral-range-bound}
\end{equation}

For the chirality operator,
Eq.~\eqref{eq:S1-kappa-spectrum} gives
\begin{equation}
\lambda_{\min}(\kappa_{i,j})
=
-\frac12,
\qquad
\lambda_{\max}(\kappa_{i,j})
=
+\frac12.
\end{equation}
Hence
\begin{align}
\operatorname{Var}_{\rho}(\kappa_{i,j})
&\le
\frac14
\left[
\frac12-\left(-\frac12\right)
\right]^2
\notag\\
&=
\frac14.
\label{eq:S1-single-bond-bound}
\end{align}

The upper value \(1/4\) is attainable.
For example, in the state
\(|\uparrow\downarrow\rangle\),
\begin{equation}
\langle\kappa_{i,j}\rangle=0,
\qquad
\langle\kappa_{i,j}^2\rangle=\frac14,
\end{equation}
and therefore
\begin{equation}
\operatorname{Var}_{\uparrow\downarrow}
(\kappa_{i,j})
=
\frac14.
\end{equation}
Equivalently, defining the chirality eigenstates
\begin{equation}
|\chi_+\rangle
=
\frac{
|\uparrow\downarrow\rangle
-i|\downarrow\uparrow\rangle
}{\sqrt2},
\qquad
|\chi_-\rangle
=
\frac{
|\uparrow\downarrow\rangle
+i|\downarrow\uparrow\rangle
}{\sqrt2},
\label{eq:S1-chirality-eigenstates}
\end{equation}
which satisfy
\begin{equation}
\kappa_{i,j}|\chi_\pm\rangle
=
\pm\frac12|\chi_\pm\rangle,
\end{equation}
one has
\begin{equation}
|\uparrow\downarrow\rangle
=
\frac{
|\chi_+\rangle+|\chi_-\rangle
}{\sqrt2}.
\end{equation}
Thus the largest and smallest chirality eigenvalues occur with
equal probability in this state, which realizes the maximal
spectral-range variance.
}

We will also repeatedly use the covariance Cauchy--Schwarz
inequality.
For a density matrix \(\rho\), define
\begin{equation}
(X,Y)_\rho
=
\operatorname{Tr}(\rho X^\dagger Y).
\label{eq:S1-rho-inner}
\end{equation}
The Cauchy--Schwarz inequality gives
\begin{equation}
|(X,Y)_\rho|^2
\le
(X,X)_\rho(Y,Y)_\rho.
\label{eq:S1-rho-CS}
\end{equation}
{\color{black}
For the following covariance calculation in the state \(\rho\),
the fluctuation operators are centered with respect to this
same density matrix:
\(\Delta A=A-\operatorname{Tr}(\rho A)I\) and
\(\Delta B=B-\operatorname{Tr}(\rho B)I\).
}
Taking
\(X=\Delta A\) and \(Y=\Delta B\), and using
\begin{equation}
\operatorname{Cov}_{\rho}(A,B)
=
\operatorname{Re}
\operatorname{Tr}
\left(
\rho\Delta A\Delta B
\right),
\label{eq:S1-cov-real}
\end{equation}
we obtain
\begin{equation}
|\operatorname{Cov}_{\rho}(A,B)|
\le
\sqrt{
\operatorname{Var}_{\rho}(A)
\operatorname{Var}_{\rho}(B)
}.
\label{eq:S1-cov-CS}
\end{equation}

For two internal chirality bonds,
Eqs.~\eqref{eq:S1-single-bond-bound}
and \eqref{eq:S1-cov-CS} immediately give
\begin{equation}
|\operatorname{Cov}(\kappa_e,\kappa_f)|
\le
\frac14,
\label{eq:S1-in-in-cov}
\end{equation}
and hence the full covariance contribution in the variance
expansion satisfies
\begin{equation}
2|\operatorname{Cov}(\kappa_e,\kappa_f)|
\le
\frac12.
\label{eq:S1-in-in-full}
\end{equation}

\subsection*{S1.2 Block-product factorization}

Let \(\mathcal B(O)\) denote the set of blocks on which an
operator \(O\) acts nontrivially.
If
\begin{equation}
\mathcal B(O_1)
\cap
\mathcal B(O_2)
=
\varnothing,
\label{eq:S1-disjoint-block-support}
\end{equation}
then the product form
Eq.~\eqref{eq:S1-kprod-state} implies
\begin{equation}
\langle O_1O_2\rangle
=
\langle O_1\rangle
\langle O_2\rangle.
\label{eq:S1-factorization}
\end{equation}
Therefore
\begin{equation}
\operatorname{Cov}(O_1,O_2)=0.
\label{eq:S1-disjoint-cov-zero}
\end{equation}
Thus two matching bonds can have a nonzero covariance only
if they touch at least one common block of the chosen
\(k\)-producible partition.

\subsection*{S1.3 Effective operator generated by a cross bond}

Consider a cross bond \(e\) whose ordered chirality operator is
\begin{equation}
\kappa_e
\equiv
\kappa_{i,j}
=
S_i^xS_j^y-S_i^yS_j^x.
\label{eq:S1-cross-fixed-orientation}
\end{equation}
The ordering \((i,j)\) in
Eq.~\eqref{eq:S1-cross-fixed-orientation} is the physical
ordering inherited from the definition of the chirality
generator and will not be changed according to the block
labels.

Suppose that the endpoints belong to two distinct blocks
\(u\) and \(v\).
The effective operator induced on block \(u\) is defined,
while preserving the fixed ordering \((i,j)\), by
\begin{equation}
Y_e^{(u)}
=
\begin{cases}
\langle S_j^y\rangle_v S_i^x
-
\langle S_j^x\rangle_v S_i^y,
&
i\in u,\quad j\in v,
\\[2mm]
\langle S_i^x\rangle_v S_j^y
-
\langle S_i^y\rangle_v S_j^x,
&
i\in v,\quad j\in u.
\end{cases}
\label{eq:S1-effective-Y}
\end{equation}
The second line does not replace
\(\kappa_{i,j}\) by \(\kappa_{j,i}\); it only replaces the
spin operators belonging to block \(v\) by their one-point
expectation values.

For any operator \(X_u\) supported entirely in block \(u\),
block-product factorization gives
\begin{align}
\langle X_u\kappa_e\rangle
&=
\langle X_uY_e^{(u)}\rangle_u,
\label{eq:S1-effective-left}
\\
\langle \kappa_eX_u\rangle
&=
\langle Y_e^{(u)}X_u\rangle_u,
\label{eq:S1-effective-right}
\\
\langle\kappa_e\rangle
&=
\langle Y_e^{(u)}\rangle_u.
\label{eq:S1-effective-mean}
\end{align}

Let \(\ell\) denote the endpoint of \(e\) in block \(u\) and
\(m\) the endpoint in block \(v\).
In either line of Eq.~\eqref{eq:S1-effective-Y},
\(Y_e^{(u)}\) has the form
\begin{equation}
Y_e^{(u)}
=
a_mS_\ell^x+b_mS_\ell^y,
\label{eq:S1-Y-linear}
\end{equation}
with
\begin{equation}
a_m^2+b_m^2
=
\langle S_m^x\rangle_v^2
+
\langle S_m^y\rangle_v^2.
\label{eq:S1-Y-coeff}
\end{equation}
For a single spin-\(1/2\),
\begin{equation}
\sqrt{
\langle S_m^x\rangle_v^2+
\langle S_m^y\rangle_v^2
}
\le
\frac12.
\label{eq:S1-mperp}
\end{equation}
{\color{black}
Indeed, {\color{black}when the transverse magnetization is nonzero,}
choosing an in-plane spin direction
\(S_m^{\textcolor{black}{\theta}}\)
parallel to the vector
\((\langle S_m^x\rangle_v,\langle S_m^y\rangle_v)\), one has
\begin{equation}
\sqrt{
\langle S_m^x\rangle_v^2+
\langle S_m^y\rangle_v^2
}
=
\left|\langle S_m^{\textcolor{black}{\theta}}\rangle_v\right|
\le
\|S_m^{\textcolor{black}{\theta}}\|
=
\frac12,
\label{eq:S1-mperp-proof}
\end{equation}
where the last equality follows because any spin-\(1/2\)
component has eigenvalues \(\pm1/2\).
}
{\color{black}
Here
\(m_m^\perp=\sqrt{\langle S_m^x\rangle_v^2+
\langle S_m^y\rangle_v^2}\).
If \(m_m^\perp=0\), then \(Y_e^{(u)}=0\) and
\(\operatorname{Var}_u(Y_e^{(u)})=0\);
the angle \(\phi\) below may then be chosen arbitrarily.
}
Hence, for some in-plane spin component \(S_\ell^\phi\),
\begin{equation}
Y_e^{(u)}
=
m_m^\perp S_\ell^\phi,
\qquad
0\le m_m^\perp\le\frac12.
\label{eq:S1-Y-spin-component}
\end{equation}
Since the eigenvalues of any spin-\(1/2\) component are
\(\pm1/2\),
\begin{equation}
\operatorname{Var}_u(S_\ell^\phi)
\le
\frac14.
\label{eq:S1-spin-component-var}
\end{equation}
Therefore
\begin{equation}
\operatorname{Var}_u(Y_e^{(u)})
\le
\frac1{16}.
\label{eq:S1-effective-var}
\end{equation}

\subsection*{S1.4 Internal--cross covariance}

Let \(\kappa_e^{\rm in}\) be an internal matching bond of
block \(u\), and let \(\kappa_f^{\rm cr}\) be a cross matching
bond incident on \(u\).

{\color{black}
Starting from the symmetrized covariance,
\begin{align}
\operatorname{Cov}_{\Psi^{(k)}}
\left(
\kappa_e^{\rm in},
\kappa_f^{\rm cr}
\right)
&=
\frac12
\left[
\left\langle
\kappa_e^{\rm in}\kappa_f^{\rm cr}
\right\rangle
+
\left\langle
\kappa_f^{\rm cr}\kappa_e^{\rm in}
\right\rangle
\right]
\notag\\
&\quad
-
\left\langle\kappa_e^{\rm in}\right\rangle
\left\langle\kappa_f^{\rm cr}\right\rangle .
\label{eq:S1-in-cross-expanded}
\end{align}
Because \(\kappa_e^{\rm in}\) acts entirely within block
\(u\),
\begin{equation}
\left\langle\kappa_e^{\rm in}\right\rangle
=
\left\langle\kappa_e^{\rm in}\right\rangle_u .
\end{equation}
Using
Eqs.~\eqref{eq:S1-effective-left}--%
\eqref{eq:S1-effective-mean} with
\(X_u=\kappa_e^{\rm in}\), we have
\begin{align}
\left\langle
\kappa_e^{\rm in}\kappa_f^{\rm cr}
\right\rangle
&=
\left\langle
\kappa_e^{\rm in}Y_f^{(u)}
\right\rangle_u,
\\
\left\langle
\kappa_f^{\rm cr}\kappa_e^{\rm in}
\right\rangle
&=
\left\langle
Y_f^{(u)}\kappa_e^{\rm in}
\right\rangle_u,
\\
\left\langle
\kappa_f^{\rm cr}
\right\rangle
&=
\left\langle
Y_f^{(u)}
\right\rangle_u.
\end{align}
Substitution into
Eq.~\eqref{eq:S1-in-cross-expanded} gives
}
\begin{equation}
\operatorname{Cov}_{\Psi^{(k)}}
\left(
\kappa_e^{\rm in},
\kappa_f^{\rm cr}
\right)
=
\operatorname{Cov}_{u}
\left(
\kappa_e^{\rm in},
Y_f^{(u)}
\right).
\label{eq:S1-in-cross-reduction}
\end{equation}

Equations~\eqref{eq:S1-single-bond-bound},
\eqref{eq:S1-effective-var}, and
\eqref{eq:S1-cov-CS} then yield
\begin{align}
\left|
\operatorname{Cov}_{\Psi^{(k)}}
\left(
\kappa_e^{\rm in},
\kappa_f^{\rm cr}
\right)
\right|
&\le
\sqrt{
\frac14\frac1{16}
}
\notag\\
&=
\frac18.
\label{eq:S1-in-cross-bound}
\end{align}
Thus
\begin{equation}
2\left|
\operatorname{Cov}_{\Psi^{(k)}}
\left(
\kappa_e^{\rm in},
\kappa_f^{\rm cr}
\right)
\right|
\le
\frac14.
\label{eq:S1-in-cross-full}
\end{equation}

\subsection*{S1.5 Cross--cross covariance with one shared block}

Consider two cross matching bonds \(e\) and \(f\) that share
exactly one block \(u\).
Denote their endpoints on \(u\) by \(A\) and \(C\), and their
other endpoints by \(B\in v\) and \(D\in w\), respectively:
\begin{equation}
A,C\in u,
\qquad
B\in v,
\qquad
D\in w,
\qquad
u,v,w\ {\rm distinct}.
\label{eq:S1-one-shared-geometry}
\end{equation}
Here \(A,B,C,D\) label the physical endpoints according to
their block membership; they do not redefine the fixed
ordering of either chirality operator.

{\color{black}
In particular, depending on the fixed physical orientation
inherited from \(K_{\rm tot}\), the first bond is either
\begin{equation}
\kappa_e=\kappa_{A,B}
\qquad\text{or}\qquad
\kappa_e=\kappa_{B,A},
\end{equation}
and similarly
\begin{equation}
\kappa_f=\kappa_{C,D}
\qquad\text{or}\qquad
\kappa_f=\kappa_{D,C}.
\end{equation}
These alternatives do not represent a reorientation of the
bonds; they only describe the possible relation between the
fixed chirality ordering and the block-based endpoint labels.

To show explicitly how the block-product factorization works,
consider first the representative case
\begin{equation}
\kappa_e=\kappa_{A,B},
\qquad
\kappa_f=\kappa_{C,D}.
\end{equation}
Then
\begin{align}
\kappa_e\kappa_f
&=
\left(
S_A^xS_B^y-S_A^yS_B^x
\right)
\left(
S_C^xS_D^y-S_C^yS_D^x
\right)
\notag\\
&=
S_A^xS_B^yS_C^xS_D^y
-
S_A^xS_B^yS_C^yS_D^x
\notag\\
&\quad
-
S_A^yS_B^xS_C^xS_D^y
+
S_A^yS_B^xS_C^yS_D^x.
\label{eq:S1-one-shared-product-expansion}
\end{align}
Since \(A,C\in u\), \(B\in v\), and \(D\in w\),
and \(u,v,w\) are mutually distinct, the pure block-product
state gives
\begin{align}
\langle\kappa_e\kappa_f\rangle
&=
\langle S_A^xS_C^x\rangle_u
\langle S_B^y\rangle_v
\langle S_D^y\rangle_w
\notag\\
&\quad
-
\langle S_A^xS_C^y\rangle_u
\langle S_B^y\rangle_v
\langle S_D^x\rangle_w
\notag\\
&\quad
-
\langle S_A^yS_C^x\rangle_u
\langle S_B^x\rangle_v
\langle S_D^y\rangle_w
\notag\\
&\quad
+
\langle S_A^yS_C^y\rangle_u
\langle S_B^x\rangle_v
\langle S_D^x\rangle_w.
\label{eq:S1-one-shared-factorized-product}
\end{align}
For this orientation,
\begin{align}
Y_e^{(u)}
&=
\langle S_B^y\rangle_v S_A^x
-
\langle S_B^x\rangle_v S_A^y,
\\
Y_f^{(u)}
&=
\langle S_D^y\rangle_w S_C^x
-
\langle S_D^x\rangle_w S_C^y.
\end{align}
Their product is
\begin{align}
Y_e^{(u)}Y_f^{(u)}
&=
\langle S_B^y\rangle_v
\langle S_D^y\rangle_w
S_A^xS_C^x
\notag\\
&\quad
-
\langle S_B^y\rangle_v
\langle S_D^x\rangle_w
S_A^xS_C^y
\notag\\
&\quad
-
\langle S_B^x\rangle_v
\langle S_D^y\rangle_w
S_A^yS_C^x
\notag\\
&\quad
+
\langle S_B^x\rangle_v
\langle S_D^x\rangle_w
S_A^yS_C^y.
\end{align}
Therefore
\begin{equation}
\langle\kappa_e\kappa_f\rangle
=
\left\langle
Y_e^{(u)}Y_f^{(u)}
\right\rangle_u.
\label{eq:S1-one-shared-forward}
\end{equation}
Repeating the same factorization with the reversed operator
order gives
\begin{equation}
\langle\kappa_f\kappa_e\rangle
=
\left\langle
Y_f^{(u)}Y_e^{(u)}
\right\rangle_u.
\label{eq:S1-one-shared-reverse}
\end{equation}
The remaining three possible fixed index orderings,
\begin{equation}
(\kappa_{A,B},\kappa_{D,C}),
\qquad
(\kappa_{B,A},\kappa_{C,D}),
\qquad
(\kappa_{B,A},\kappa_{D,C}),
\end{equation}
are treated in exactly the same way.
The signs and the placement of the \(x\) and \(y\) spin
components agree with the corresponding case in
Eq.~\eqref{eq:S1-effective-Y}.
Hence
Eqs.~\eqref{eq:S1-one-shared-forward} and
\eqref{eq:S1-one-shared-reverse} hold for all fixed
orientations.

Together with
\begin{equation}
\langle\kappa_e\rangle
=
\langle Y_e^{(u)}\rangle_u,
\qquad
\langle\kappa_f\rangle
=
\langle Y_f^{(u)}\rangle_u,
\end{equation}
the definition of the symmetrized covariance therefore gives
}
\begin{equation}
\operatorname{Cov}_{\Psi^{(k)}}
\left(
\kappa_e^{\rm cr},
\kappa_f^{\rm cr}
\right)
=
\operatorname{Cov}_{u}
\left(
Y_e^{(u)},
Y_f^{(u)}
\right).
\label{eq:S1-one-shared-reduction}
\end{equation}

Using Eq.~\eqref{eq:S1-effective-var} for both effective
operators and then Eq.~\eqref{eq:S1-cov-CS},
\begin{align}
\left|
\operatorname{Cov}_{\Psi^{(k)}}
\left(
\kappa_e^{\rm cr},
\kappa_f^{\rm cr}
\right)
\right|
&\le
\sqrt{
\frac1{16}\frac1{16}
}
\notag\\
&=
\frac1{16}.
\label{eq:S1-one-shared-bound}
\end{align}
Therefore
\begin{equation}
2\left|
\operatorname{Cov}_{\Psi^{(k)}}
\left(
\kappa_e^{\rm cr},
\kappa_f^{\rm cr}
\right)
\right|
\le
\frac18.
\label{eq:S1-one-shared-full}
\end{equation}

\subsection*{S1.6 Cross--cross covariance with the same two shared blocks}

We now consider the only geometry that cannot be reduced to
two independent one-point functions on the outer blocks.
Let \(e\) and \(f\) be two distinct matching bonds connecting
the same two blocks \(u\) and \(v\).
Denote their endpoints according to block membership by
\begin{equation}
A,C\in u,
\qquad
B,D\in v.
\label{eq:S1-same-two-geometry}
\end{equation}
Again, these labels do not alter the fixed physical ordering
of the chirality operators.
Because \(e\) and \(f\) are matching bonds, all four physical
sites are distinct and
\begin{equation}
[\kappa_e,\kappa_f]=0.
\label{eq:S1-same-two-commute}
\end{equation}

Define
\begin{equation}
\begin{aligned}
V_e
&=
\operatorname{Var}_{\Psi^{(k)}}(\kappa_e),
\qquad
V_f
=
\operatorname{Var}_{\Psi^{(k)}}(\kappa_f),
\\
C_{ef}
&=
\operatorname{Cov}_{\Psi^{(k)}}(\kappa_e,\kappa_f).
\end{aligned}
\label{eq:S1-VeVfC}
\end{equation}
The symmetrized covariance is real.
If \(C_{ef}=0\), the desired bound is immediate.
Otherwise choose
\(\eta\in\{+1,-1\}\) such that
\begin{equation}
\eta C_{ef}=|C_{ef}|
\label{eq:S1-eta-choice}
\end{equation}
and define
\begin{equation}
L_\eta
=
\kappa_e+\eta\kappa_f.
\label{eq:S1-Leta}
\end{equation}
Then
\begin{equation}
\operatorname{Var}(L_\eta)
=
V_e+V_f+2|C_{ef}|.
\label{eq:S1-Var-Leta}
\end{equation}

After tracing out all other sites, the four-site reduced state
factorizes between the two \(k\)-producible blocks,
\begin{equation}
\rho_{ACBD}
=
\rho_{AC}\otimes\rho_{BD}.
\label{eq:S1-four-site-product}
\end{equation}
The reduced states \(\rho_{AC}\) and \(\rho_{BD}\) may
themselves be entangled.

Define the following operator on the \(BD\) Hilbert space:
\begin{equation}
M_\eta
=
\operatorname{Tr}_{AC}
\left[
(\sqrt{\rho_{AC}}\otimes I_{BD})
L_\eta^2
(\sqrt{\rho_{AC}}\otimes I_{BD})
\right].
\label{eq:S1-Meta}
\end{equation}
The operator inside the partial trace is positive
semidefinite, and partial trace preserves positivity.
Therefore
\begin{equation}
M_\eta\ge0.
\label{eq:S1-Meta-positive}
\end{equation}
Moreover,
\begin{equation}
\operatorname{Tr}_{BD}
\left(
\rho_{BD}M_\eta
\right)
=
\langle L_\eta^2\rangle.
\label{eq:S1-Meta-expectation}
\end{equation}

{\color{black}
We now determine the structure of \(M_\eta\) explicitly.
Choose the \(BD\) basis in the order
\begin{equation}
|\uparrow_B\uparrow_D\rangle,
\quad
|\downarrow_B\downarrow_D\rangle,
\quad
|\uparrow_B\downarrow_D\rangle,
\quad
|\downarrow_B\uparrow_D\rangle.
\label{eq:S1-BD-basis}
\end{equation}
The first two states span
\begin{equation}
\mathcal H_+
=
\operatorname{span}
\left\{
|\uparrow_B\uparrow_D\rangle,
|\downarrow_B\downarrow_D\rangle
\right\},
\end{equation}
whereas the last two span
\begin{equation}
\mathcal H_-
=
\operatorname{span}
\left\{
|\uparrow_B\downarrow_D\rangle,
|\downarrow_B\uparrow_D\rangle
\right\}.
\end{equation}

The terms \(\kappa_e^2\) and \(\kappa_f^2\) do not change
the \(S^z\) configuration of sites \(B,D\).
In contrast, \(\kappa_e\kappa_f\) flips both \(B\) and \(D\)
once.
Since
\begin{equation}
L_\eta^2
=
\kappa_e^2+\kappa_f^2
+
2\eta\kappa_e\kappa_f,
\label{eq:S1-Leta-square}
\end{equation}
\(M_\eta\) is block diagonal with respect to
\(\mathcal H_+\oplus\mathcal H_-\).

Define
\begin{equation}
z_A
=
\operatorname{Tr}(\rho_{AC}S_A^z),
\qquad
z_C
=
\operatorname{Tr}(\rho_{AC}S_C^z).
\label{eq:S1-zA-zC}
\end{equation}
For \(s_B,s_D\in\{\uparrow,\downarrow\}\), let
\begin{equation}
d_{s_Bs_D}
=
\langle s_Bs_D|M_\eta|s_Bs_D\rangle,
\end{equation}
and write
\begin{equation}
S^z|s\rangle
=
m_s|s\rangle,
\qquad
m_\uparrow=\frac12,
\qquad
m_\downarrow=-\frac12.
\end{equation}
From the sandwich definition
Eq.~\eqref{eq:S1-Meta},
\begin{align}
d_{s_Bs_D}
&=
\operatorname{Tr}_{AC}
\left[
\sqrt{\rho_{AC}}\,
\langle s_Bs_D|L_\eta^2|s_Bs_D\rangle
\sqrt{\rho_{AC}}
\right]
\notag\\
&=
\operatorname{Tr}_{AC}
\left[
\rho_{AC}\,
\langle s_Bs_D|L_\eta^2|s_Bs_D\rangle
\right].
\label{eq:S1-Meta-diag-partial-trace}
\end{align}

{\color{black}
The identity in Eq.~\eqref{eq:S1-kappa-square} is independent
of the orientation of the ordered bond.
Indeed,
\(\kappa_{j,i}=-\kappa_{i,j}\), and therefore
\begin{equation}
\kappa_{j,i}^2
=
\kappa_{i,j}^2
=
\frac18 I-\frac12S_i^zS_j^z.
\label{eq:S1-kappa-square-orientation}
\end{equation}
Hence the following evaluation is valid irrespective of
whether the fixed physical ordering of bond \(e\) is
\((A,B)\) or \((B,A)\), and likewise irrespective of whether
the ordering of \(f\) is \((C,D)\) or \((D,C)\).
}
Using
Eq.~\eqref{eq:S1-kappa-square}, we have
\begin{align}
\langle s_B{\color{black}s_D}|
\kappa_e^2
|s_B{\color{black}s_D}\rangle
&=
\frac18 I_{AC}
-
\frac12m_{s_B}S_A^z,
\\
\langle {\color{black}s_B}s_D|
\kappa_f^2
|{\color{black}s_B}s_D\rangle
&=
\frac18 I_{AC}
-
\frac12m_{s_D}S_C^z.
\end{align}
The cross term
\(\kappa_e\kappa_f\) contains \(S_B^x\) or \(S_B^y\)
and \(S_D^x\) or \(S_D^y\).
Since
\begin{equation}
\langle s|S^x|s\rangle
=
\langle s|S^y|s\rangle
=
0
\end{equation}
in the \(S^z\) basis,
\begin{equation}
\langle s_Bs_D|
\kappa_e\kappa_f
|s_Bs_D\rangle
=
0.
\end{equation}
Consequently,
\begin{equation}
d_{s_Bs_D}
=
\frac14
-
\frac12m_{s_B}z_A
-
\frac12m_{s_D}z_C.
\label{eq:S1-Meta-diag-general}
\end{equation}
The four diagonal entries are therefore
\begin{align}
d_{\uparrow\uparrow}
&=
\frac14-\frac14z_A-\frac14z_C,
&
d_{\uparrow\downarrow}
&=
\frac14-\frac14z_A+\frac14z_C,
\label{eq:S1-Meta-diag-1}
\\
d_{\downarrow\uparrow}
&=
\frac14+\frac14z_A-\frac14z_C,
&
d_{\downarrow\downarrow}
&=
\frac14+\frac14z_A+\frac14z_C.
\label{eq:S1-Meta-diag-2}
\end{align}

In the basis of
Eq.~\eqref{eq:S1-BD-basis}, the matrix therefore has the form
\begin{equation}
M_\eta
=
\begin{pmatrix}
d_{\uparrow\uparrow} & x_+ & 0 & 0\\
x_+^* & d_{\downarrow\downarrow} & 0 & 0\\
0 & 0 & d_{\uparrow\downarrow} & x_-\\
0 & 0 & x_-^* & d_{\downarrow\uparrow}
\end{pmatrix},
\label{eq:S1-Meta-matrix}
\end{equation}
where \(x_+\) and \(x_-\) arise from the cross term
\(2\eta\kappa_e\kappa_f\).
Their explicit values are not required for the bound.

The traces of the two parity blocks are
\begin{align}
\operatorname{Tr}_{\mathcal H_+}M_\eta
&=
d_{\uparrow\uparrow}
+
d_{\downarrow\downarrow}
\notag\\
&=
\left(
\frac14-\frac14z_A-\frac14z_C
\right)
+
\left(
\frac14+\frac14z_A+\frac14z_C
\right)
\notag\\
&=
\frac12,
\label{eq:S1-Hplus-trace}
\end{align}
and
\begin{align}
\operatorname{Tr}_{\mathcal H_-}M_\eta
&=
d_{\uparrow\downarrow}
+
d_{\downarrow\uparrow}
\notag\\
&=
\left(
\frac14-\frac14z_A+\frac14z_C
\right)
+
\left(
\frac14+\frac14z_A-\frac14z_C
\right)
\notag\\
&=
\frac12.
\label{eq:S1-Hminus-trace}
\end{align}

Because \(M_\eta\ge0\), both \(2\times2\) parity blocks
are positive semidefinite.
All eigenvalues of each block are therefore nonnegative.
For a positive semidefinite matrix the largest eigenvalue
cannot exceed its trace, and each parity-block trace is
\(1/2\).
Hence every eigenvalue of \(M_\eta\) is at most \(1/2\), so
\begin{equation}
0
\le
M_\eta
\le
\frac12 I_{BD}.
\label{eq:S1-Meta-half}
\end{equation}
}

Combining Eqs.~\eqref{eq:S1-Meta-expectation} and
\eqref{eq:S1-Meta-half},
\begin{equation}
\langle L_\eta^2\rangle
\le
\frac12.
\label{eq:S1-Leta-second-half}
\end{equation}
Therefore
\begin{equation}
\operatorname{Var}(L_\eta)
\le
\frac12.
\label{eq:S1-Leta-var-half}
\end{equation}
Together with Eq.~\eqref{eq:S1-Var-Leta}, this gives the joint
inequality
\begin{equation}
\frac{V_e+V_f}{2}
+
|C_{ef}|
\le
\frac14.
\label{eq:S1-same-two-joint}
\end{equation}

To obtain the pairwise covariance bound, the covariance
Cauchy--Schwarz inequality and the arithmetic--geometric mean
inequality give
\begin{equation}
|C_{ef}|
\le
\sqrt{V_eV_f}
\le
\frac{V_e+V_f}{2}.
\label{eq:S1-same-two-CS}
\end{equation}
Hence
\begin{equation}
V_e+V_f
\ge
2|C_{ef}|.
\label{eq:S1-same-two-V}
\end{equation}
Equation~\eqref{eq:S1-Var-Leta} then implies
\begin{equation}
\operatorname{Var}(L_\eta)
\ge
4|C_{ef}|.
\label{eq:S1-Leta-lower}
\end{equation}
Using Eq.~\eqref{eq:S1-Leta-var-half}, we obtain
\begin{equation}
|C_{ef}|
\le
\frac18,
\label{eq:S1-same-two-pairwise}
\end{equation}
or equivalently
\begin{equation}
2|\operatorname{Cov}(\kappa_e,\kappa_f)|
\le
\frac14.
\label{eq:S1-same-two-full}
\end{equation}

The sandwich form in Eq.~\eqref{eq:S1-Meta} is useful because
it makes the positivity of \(M_\eta\) explicit.
Writing only
\(\operatorname{Tr}_{AC}
[(\rho_{AC}\otimes I)L_\eta^2]\)
would give the same expectation value after the final trace,
but positivity is not manifest from that representation
because the factors need not commute.

\subsection*{S1.7 Block-wise allowance}

We now return to the variance of one complete matching.
For the fixed state and partition,
\begin{align}
\operatorname{Var}_{\Psi^{(k)}}(K_{\mathcal M})
&=
\sum_{e\in\mathcal M}
\operatorname{Var}_{\Psi^{(k)}}(\kappa_e)
\notag\\
&\quad+
2
\sum_{\substack{e<f\\e,f\in\mathcal M}}
\operatorname{Cov}_{\Psi^{(k)}}(\kappa_e,\kappa_f)
\notag\\
&\le
\sum_{e\in\mathcal M}
\operatorname{Var}_{\Psi^{(k)}}(\kappa_e)
\notag\\
&\quad+
2
\sum_{\substack{e<f\\e,f\in\mathcal M}}
\left|
\operatorname{Cov}_{\Psi^{(k)}}(\kappa_e,\kappa_f)
\right|.
\label{eq:S1-KM-absolute}
\end{align}
Pairs touching disjoint sets of blocks vanish by
Eq.~\eqref{eq:S1-disjoint-cov-zero}.
The remaining contributions can be assigned block by block.

For each \(u\in\Gamma\), define
{\color{black}
\begin{equation}
A_u
=
\frac{r_u}{4}
+
\frac{c_u}{8}
+
2\binom{r_u}{2}\frac14
+
2r_uc_u\frac18
+
\binom{c_u}{2}\frac18.
\label{eq:S1-Au}
\end{equation}
}
The five terms have the following origin.

For the \(r_u\) internal bonds, the single-bond variance bound
gives the contribution
\begin{equation}
\frac{r_u}{4}.
\label{eq:S1-Au-in-var}
\end{equation}
Each cross bond has variance at most \(1/4\) and touches two
blocks.
Splitting this contribution equally between the two endpoint
blocks gives
\begin{equation}
\frac{c_u}{8}
\label{eq:S1-Au-cross-var}
\end{equation}
for block \(u\).

There are
\(\binom{r_u}{2}\) unordered internal--internal pairs.
Equation~\eqref{eq:S1-in-in-full} therefore gives
\begin{equation}
2\binom{r_u}{2}\frac14
=
\frac12\binom{r_u}{2}.
\label{eq:S1-Au-in-in}
\end{equation}
There are \(r_uc_u\) internal--cross pairs incident on \(u\).
Equation~\eqref{eq:S1-in-cross-full} gives
\begin{equation}
2r_uc_u\frac18
=
\frac14r_uc_u.
\label{eq:S1-Au-in-cross}
\end{equation}

Finally, the \(c_u\) cross bonds incident on \(u\) form
\begin{equation}
\binom{c_u}{2}
\label{eq:S1-cross-pair-number}
\end{equation}
unordered pairs.
These pairs are partitioned into the two cross--cross
geometries shown in Fig.~\ref{fig:chirality-bound-structure}d.

If a pair shares only block \(u\),
Eq.~\eqref{eq:S1-one-shared-full} shows that its full
contribution is at most \(1/8\), and that pair appears in the
allowance of \(u\) only once.

If instead the same pair connects the same two blocks \(u\)
and \(v\), Eq.~\eqref{eq:S1-same-two-full} gives a full
contribution at most \(1/4\).
The physical pair is then counted once among the
\(\binom{c_u}{2}\) pairs incident on \(u\) and once among the
\(\binom{c_v}{2}\) pairs incident on \(v\).
Assigning \(1/8\) to each shared block therefore provides
\begin{equation}
\frac18+\frac18=\frac14,
\label{eq:S1-same-two-allocation}
\end{equation}
which covers the full covariance contribution.

{\color{black}
It is important that
\(\binom{c_u}{2}\) denotes the total number of unordered pairs
of cross bonds incident on block \(u\).
It is not the number of one-shared-block pairs plus another
independent \(\binom{c_u}{2}\) same-two-block pairs.
Each unordered pair belongs to exactly one of these two
geometrical cases.
Thus the single contribution
\begin{equation}
\binom{c_u}{2}\frac18
\end{equation}
covers all cross--cross pairs incident on \(u\), with the
same-two-block case receiving the second \(1/8\) from the
other shared block.
}

Consequently, assigning \(1/8\) per unordered cross--cross pair
incident on each block covers both geometries, giving the last
term in Eq.~\eqref{eq:S1-Au}.
Every term in Eq.~\eqref{eq:S1-KM-absolute} is therefore
covered by the block allowances, and hence
\begin{equation}
\operatorname{Var}_{\Psi^{(k)}}(K_{\mathcal M})
\le
\sum_{u\in\Gamma}A_u.
\label{eq:S1-var-sum-Au}
\end{equation}

The allowance can be simplified algebraically:
\begin{align}
A_u
&=
\frac{r_u}{4}
+
\frac{c_u}{8}
+
\frac{r_u(r_u-1)}{4}
+
\frac{r_uc_u}{4}
+
\frac{c_u(c_u-1)}{16}
\notag\\
&=
\frac{r_u^2}{4}
+
\frac{r_uc_u}{4}
+
\frac{c_u^2+c_u}{16}
\notag\\
&=
\frac{(2r_u+c_u)^2+c_u}{16}
\notag\\
&=
\frac{n_u^2+c_u}{16}
\le
\frac{n_u^2+n_u}{16},
\label{eq:S1-Au-nu}
\end{align}
where Eq.~\eqref{eq:S1-coverage} and
\(0\le c_u\le n_u\) were used.
Thus
\begin{equation}
\operatorname{Var}_{\Psi^{(k)}}(K_{\mathcal M})
\le
\frac1{16}
\sum_{u\in\Gamma}
\left(
n_u^2+n_u
\right).
\label{eq:S1-block-dependent-bound}
\end{equation}

{\color{black}
The remaining maximization over all admissible block
partitions under the \(k\)-producibility constraint is carried
out in Step 5 of Sec.~II.
The subsequent recombination of the odd and even matching
bounds using the triangle inequality for the square root of
the QFI is carried out in Step 6 of Sec.~II, with the
triangle-type inequality itself proved in Supplementary Note S2.
}

{\color{black}

\section*{Supplementary Note S2:
Proof of the triangle-type inequality for QFI}

Here we give an alternative proof of the inequality
\begin{equation}
\sqrt{F_Q[\rho,A+B]}
\le
\sqrt{F_Q[\rho,A]}
+
\sqrt{F_Q[\rho,B]},
\label{eq:S1alt-triangle}
\end{equation}
which does not require the symmetric logarithmic derivative.
Let
\begin{equation}
\rho
=
\sum_m p_m |m\rangle\langle m|
\label{eq:S1alt-spectral-rho}
\end{equation}
be the spectral decomposition of the density matrix.
For a unitary parametrization generated by a Hermitian operator
\(G\), the quantum Fisher information can be written in the
spectral form~\cite{Toth2014,Liu2020}
\begin{equation}
F_Q[\rho,G]
=
2
\sum_{\substack{m,n\\p_m+p_n>0}}
\frac{(p_m-p_n)^2}{p_m+p_n}
\left|
\langle m|G|n\rangle
\right|^2 .
\label{eq:S1alt-spectral-QFI}
\end{equation}
The restriction \(p_m+p_n>0\) makes
Eq.~\eqref{eq:S1alt-spectral-QFI} directly applicable also to
rank-deficient density matrices.

For fixed \(\rho\), define
\begin{equation}
X_{mn}(G)
=
\left[
2\frac{(p_m-p_n)^2}{p_m+p_n}
\right]^{1/2}
\langle m|G|n\rangle ,
\qquad
p_m+p_n>0 .
\label{eq:S1alt-X-def}
\end{equation}
Equation~\eqref{eq:S1alt-spectral-QFI} then becomes
\begin{equation}
F_Q[\rho,G]
=
\sum_{\substack{m,n\\p_m+p_n>0}}
|X_{mn}(G)|^2,
\label{eq:S1alt-QFI-norm}
\end{equation}
or equivalently
\begin{equation}
\sqrt{F_Q[\rho,G]}
=
\|X(G)\|_2 ,
\label{eq:S1alt-sqrt-QFI-norm}
\end{equation}
where \(\|\cdot\|_2\) denotes the Euclidean norm of the collection
of matrix elements \(X_{mn}\).

Since the matrix elements of the generator are linear in \(G\),
\begin{equation}
X(A+B)
=
X(A)+X(B).
\label{eq:S1alt-linearity}
\end{equation}
The ordinary triangle inequality for the Euclidean norm therefore
gives
\begin{align}
\sqrt{F_Q[\rho,A+B]}
&=
\|X(A+B)\|_2
\nonumber\\
&=
\|X(A)+X(B)\|_2
\nonumber\\
&\le
\|X(A)\|_2+\|X(B)\|_2
\nonumber\\
&=
\sqrt{F_Q[\rho,A]}
+
\sqrt{F_Q[\rho,B]} ,
\label{eq:S1alt-proof}
\end{align}
which proves Eq.~\eqref{eq:S1alt-triangle}.

}


\section*{Supplementary Note S3: TPQ calculation of the finite-temperature chirality QFI}

The finite-temperature chirality QFI presented in Fig.~\ref{fig:qfi-tpq-idmrg}
was evaluated using canonical thermal pure quantum (TPQ)
states~\cite{Imada1986,Hams2000,Sugiura2013,Ikeuchi2015,Endo2018,Tokuro2025}.
\textcolor{black}{The target quantity is the QFI of the thermal Gibbs
state, estimated through TPQ correlation functions, rather than the
pure-state QFI of an individual TPQ realization.}
The calculations were carried out for periodic spin-\(1/2\) chains with
\(N=12\), \textcolor{black}{\(16\),} \(20\), and \(28\) sites in the full Hilbert space of
dimension
\begin{equation}
D=2^N.
\label{eq:S2-Hilbert-dimension}
\end{equation}
For a Hamiltonian \(H\), we first generate a normalized random state
\begin{equation}
|\psi_0\rangle
=
\sum_{n=1}^{D}c_n|n\rangle,
\qquad
\sum_{n=1}^{D}|c_n|^2=1,
\label{eq:S2-random-state}
\end{equation}
where the \(c_n\) are random complex coefficients and
\(\{|n\rangle\}\) is an orthonormal basis of the Hilbert space.
At inverse temperature
\begin{equation}
\beta=\frac{1}{T},
\label{eq:S2-beta}
\end{equation}
the corresponding canonical TPQ state is
\begin{equation}
|\beta,N\rangle
=
e^{-\beta H/2}|\psi_0\rangle.
\label{eq:S2-TPQ-state}
\end{equation}
For a given TPQ realization, a thermal expectation value is estimated
as
\begin{equation}
\langle A\rangle_T
\simeq
\frac{
\langle\beta,N|A|\beta,N\rangle
}{
\langle\beta,N|\beta,N\rangle
}.
\label{eq:S2-thermal-average}
\end{equation}

The imaginary-time evolution entering
Eq.~\eqref{eq:S2-TPQ-state} and the subsequent real-time evolution
were evaluated using Chebyshev-polynomial expansions. To this end, the
Hamiltonian was rescaled as
\begin{equation}
\widetilde H
=
\frac{H-b}{a},
\qquad
a=\frac{E_{\max}-E_{\min}}{2},
\qquad
b=\frac{E_{\max}+E_{\min}}{2},
\label{eq:S2-H-rescale}
\end{equation}
where \(E_{\min}\) and \(E_{\max}\) were estimated by Lanczos
calculations. The corresponding Chebyshev vectors are generated
recursively from
\begin{align}
|v_0\rangle &= |\psi\rangle,
\notag\\
|v_1\rangle &= \widetilde H|\psi\rangle,
\notag\\
|v_{k+1}\rangle
&=
2\widetilde H|v_k\rangle-|v_{k-1}\rangle .
\label{eq:S2-Cheb-vectors}
\end{align}
The real-time propagator over a time interval \(\Delta t\) can then be
written as
\begin{equation}
e^{-iH\Delta t}|\psi\rangle
=
e^{-ib\Delta t}
\left[
J_0(a\Delta t)|v_0\rangle
+
2\sum_{k=1}^{M}
(-i)^kJ_k(a\Delta t)|v_k\rangle
\right],
\label{eq:S2-Cheb-real-time}
\end{equation}
where \(J_k\) is the Bessel function of the first kind and the expansion
order \(M\) is chosen sufficiently large for convergence. The
canonical TPQ state in Eq.~\eqref{eq:S2-TPQ-state} is constructed using
the corresponding imaginary-time Chebyshev expansion, whose
coefficients are given by modified Bessel functions.

For the present calculation, the QFI generator is the total vector
chirality,
\begin{equation}
K_{\rm tot}
=
\sum_{i=1}^{N}\kappa_i^z,
\qquad
\kappa_i^z
=
(\mathbf S_i\times\mathbf S_{i+1})^z .
\label{eq:S2-Ktot}
\end{equation}
For each temperature and each TPQ realization, we construct the two
time-dependent states
\begin{align}
|A(t)\rangle
&=
e^{-iHt}|\beta,N\rangle,
\label{eq:S2-A-t}
\\
|B(t)\rangle
&=
e^{-iHt}K_{\rm tot}|\beta,N\rangle.
\label{eq:S2-B-t}
\end{align}
The real-time vector-chirality correlation function is then evaluated
as
\begin{align}
C_K(t)
&=
\frac{
\langle A(t)|K_{\rm tot}|B(t)\rangle
}{
\langle\beta,N|\beta,N\rangle
}
\notag\\
&=
\frac{
\langle\beta,N|
K_{\rm tot}(t)K_{\rm tot}(0)
|\beta,N\rangle
}{
\langle\beta,N|\beta,N\rangle
}.
\label{eq:S2-Ck-t}
\end{align}

\textcolor{black}{The TPQ correlation \(C_K(t)\) estimates the equilibrium
correlation \(C_K^{\rm eq}(t)=\langle K_{\rm tot}(t)K_{\rm tot}(0)\rangle_T\).
The latter defines the exact} dynamical vector-chirality structure
factor per spin,
\begin{equation}
S_K(\omega,T)
=
\frac{1}{2\pi N}
\int_{-\infty}^{\infty}
dt\,
e^{i\omega t}
\textcolor{black}{C_K^{\rm eq}(t)}.
\label{eq:S2-Sk-continuous}
\end{equation}
Here \(\omega\) has the same energy unit as the Hamiltonian, with
\(\textcolor{black}{\hbar=k_{\rm B}=1}\), consistently with the convention used in the main text.
\textcolor{black}{We use \(J_2\) as the reference energy scale: temperatures
and frequencies are quoted as \(T/J_2\) and \(\omega/J_2\), and the
corresponding dimensionless time is \(J_2t\).}
In practice, the real-time correlation function was evaluated on the
discrete time grid
\begin{equation}
t_j=j\,\delta t,
\qquad
\delta t=\textcolor{black}{0.1/J_2},
\qquad
0\le t_j\le t_{\max}=\textcolor{black}{300/J_2}.
\label{eq:S2-time-grid}
\end{equation}
\textcolor{black}{The number of sampled times, including \(t=0\), is
\(N_t=1+t_{\max}/\delta t=3001\).}
A Gaussian window
\begin{equation}
W(t)
=
\exp
\left[
-\frac{1}{2}
\left(
\frac{\alpha t}{t_{\max}}
\right)^2
\right],
\qquad
\alpha=5,
\label{eq:S2-window}
\end{equation}
was applied before Fourier transformation to suppress finite-time
oscillations.
\textcolor{black}{For the untruncated Gaussian window, the corresponding
frequency standard deviation is \(\sigma_\omega=\alpha/t_{\max}\), giving
\(\sigma_\omega/J_2=\alpha/(J_2t_{\max})=1/60\) for the present parameters.}
\textcolor{black}{The negative-time part was reconstructed from the
positive-time TPQ data using the equilibrium relation
\(C_K^{\rm eq}(-t)=C_K^{\rm eq}(t)^*\), which need not hold exactly for
an individual TPQ realization. The numerical spectrum was evaluated as}
{\color{black}
\begin{equation}
\begin{aligned}
&\widehat S_K^{\,\rm num}(\omega,T)
=\frac{\delta t}{2\pi N}
\biggl[{\rm Re}\,C_K(0)
\\
&\qquad
+2\,{\rm Re}\sum_{j=1}^{N_t-1}
e^{i\omega t_j}C_K(t_j)W(t_j)
\biggr].
\end{aligned}
\label{eq:S2-Sk-discrete}
\end{equation}
}
\textcolor{black}{The frequency grid was \(\omega_m=m\,\delta\omega\), with
\(m=-499,\ldots,499\) and
\(\delta\omega/J_2=2\pi/(J_2t_{\max})=2\pi/300\).}

\textcolor{black}{For the exact equilibrium spectrum}, detailed balance gives
\begin{equation}
S_K(-\omega,T)
=
e^{-\beta\omega}S_K(\omega,T).
\label{eq:S2-detailed-balance}
\end{equation}
The fluctuation--dissipation relation then connects the positive-energy
structure factor to the dissipative vector-chirality susceptibility,
\begin{equation}
\frac{\chi_{KK}''(\omega,T)}{N}
=
\pi
\left(
1-e^{-\beta\omega}
\right)
S_K(\omega,T),
\qquad
\omega>0.
\label{eq:S2-FDT}
\end{equation}
Consequently, the finite-temperature chirality-QFI density
\textcolor{black}{is expressed in terms of the exact equilibrium spectrum as}
\begin{equation}
f_Q^{(T)}(N)
=
4
\int_0^\infty
d\omega\,
\tanh\left(
\frac{\beta\omega}{2}
\right)
\left(
1-e^{-\beta\omega}
\right)
S_K(\omega,T).
\label{eq:S2-QFI-Sk}
\end{equation}
Using Eq.~\eqref{eq:S2-FDT}, this is equivalent to
\begin{equation}
f_Q^{(T)}(N)
=
\frac{4}{\pi N}
\int_0^\infty
d\omega\,
\tanh\left(
\frac{\beta\omega}{2}
\right)
\chi_{KK}''(\omega,T),
\label{eq:S2-QFI-chi}
\end{equation}
which is Eq.~\eqref{eq:fq-finite-T} of the main text.

\textcolor{black}{For each TPQ realization, the numerical estimate was
obtained by applying the positive-frequency expression in
Eq.~\eqref{eq:S2-QFI-Sk} to the numerical spectrum and replacing the
integral by the discrete sum}
{\color{black}
\begin{equation}
\widehat f_Q^{(T)}(N)
=
4\,\delta\omega
\sum_{m=0}^{499}
\tanh\left(\frac{\beta\omega_m}{2}\right)
\left(1-e^{-\beta\omega_m}\right)
\widehat S_K^{\,\rm num}(\omega_m,T).
\label{eq:S2-QFI-numerical}
\end{equation}
}
\textcolor{black}{Detailed balance and the exact fluctuation--dissipation
relation apply to \(S_K\), but are not generally preserved exactly by
the finite-time, windowed TPQ spectrum
\(\widehat S_K^{\,\rm num}\). Thus, \(\widehat f_Q^{(T)}(N)\) is a
numerical estimate of the Gibbs-state QFI density, subject to
finite-time, windowing, discretization, and TPQ sampling errors.}

Each TPQ realization was normalized and analyzed independently through
the Fourier transformation and QFI integration\textcolor{black}{.
For \(R\) independent realizations at fixed \((N,V_z/J_2,T/J_2)\),
let \(f_r\) denote the estimate in Eq.~\eqref{eq:S2-QFI-numerical}
and \(Z_r=\langle\beta,N;r|\beta,N;r\rangle\).
Since the Fourier transformation and QFI integration are linear in
the correlation function at fixed temperature, we used the ratio-of-sums
estimate}
{\color{black}
\begin{equation}
\widehat f
=\frac{\sum_{r=1}^{R}Z_r f_r}{\sum_{r=1}^{R}Z_r},
\label{eq:S3-TPQ-ratio}
\end{equation}
}
\textcolor{black}{rather than the equal-weight mean of \(f_r\).
For \(R>1\), the standard error evaluated by jackknife method was
computed from the delete-one estimates, removing the same realization
from numerator and denominator:}
{\color{black}
\begin{equation}
\begin{aligned}
\widehat f_{(-r)}
&=\frac{\sum_{s\ne r}Z_s f_s}{\sum_{s\ne r}Z_s},
\\
\mathrm{SE}_{\rm JK}
&=\left[
\frac{R-1}{R}\sum_{r=1}^{R}
\left(\widehat f_{(-r)}-\overline f_{(-)}\right)^2
\right]^{1/2},
\end{aligned}
\label{eq:S3-TPQ-jackknife}
\end{equation}
}
\textcolor{black}{where
\(\overline f_{(-)}=R^{-1}\sum_{r=1}^{R}\widehat f_{(-r)}\).
The plotted values are \(\widehat f\pm\mathrm{SE}_{\rm JK}\).
No error bar is assigned for \(R=1\).}
For the data shown in Fig.~\ref{fig:qfi-tpq-idmrg}, we used
\(40\) TPQ realizations for \(N=12\),
\textcolor{black}{\(20\) realizations for \(N=20\),
and one realization for \(N=28\)}.
\textcolor{black}{These error bars quantify TPQ sampling uncertainty and
do not include numerical errors from finite-time evolution, Gaussian
windowing, or discretization.}

\begin{figure}[!htbp]
\centering
\includegraphics[width=\columnwidth]{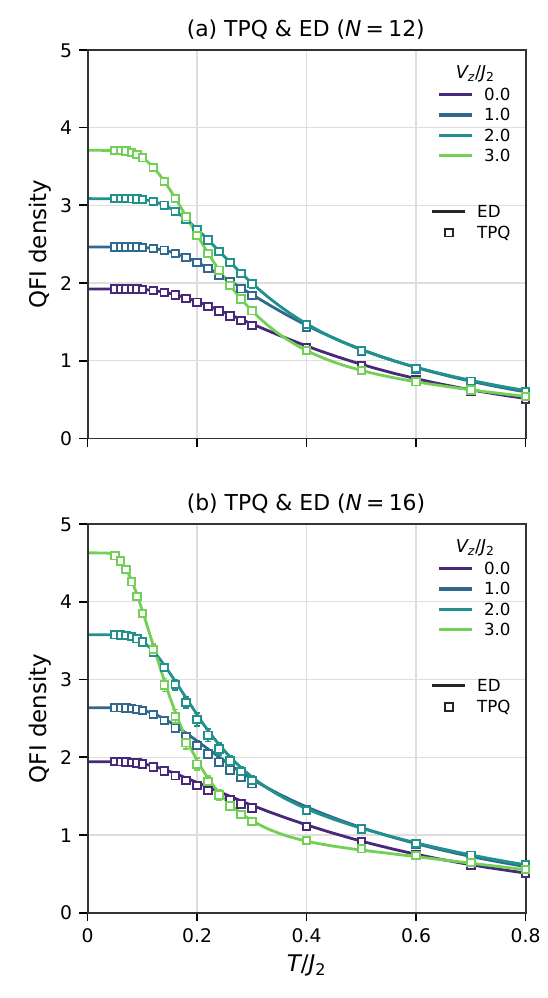}
\caption{\textcolor{black}{Temperature dependence of the chirality-QFI
density for the periodic \textcolor{black}{chains with (a) \(N=12\) and (b) \(N=16\)}, with \(J_1/J_2=-0.1\) and
\(\Delta=0.9\), at \(V_z/J_2=0.0,1.0,2.0,3.0\).
Solid lines show ED results from Eq.~\eqref{eq:S2-QFI-ED}.
Square symbols show \textcolor{black}{the ratio-of-sums TPQ estimates in
Eq.~\eqref{eq:S3-TPQ-ratio}, using 40 independent realizations for
\(N=12\) and 20 for \(N=16\);
error bars indicate one standard error evaluated by jackknife method.}
The same color denotes the same \(V_z/J_2\) for ED
and TPQ. The comparison covers \(0\le T/J_2\le0.8\).
ED includes \(T=0\), whereas the TPQ data start at
\(T/J_2=0.05\).}}
\label{fig:S3-ED-TPQ-comparison}
\end{figure}

\textcolor{black}{For the \textcolor{black}{\(N=12\) and \(16\) ED benchmarks}, we diagonalized \(H\) in
all total-\(S^z\) sectors, \textcolor{black}{resolving momentum for \(N=16\),}
retaining all \textcolor{black}{\(2^N\)} states and using
a single partition function \(Z={\rm Tr}\,e^{-\beta H}\).
The Gibbs density matrix \(\rho_T=e^{-\beta H}/Z\) was also diagonalized
numerically \textcolor{black}{at the TPQ comparison temperatures for \(N=12\)
and at selected temperatures for \(N=16\)}. Writing
\(\rho_T=\sum_a p_a|a\rangle\langle a|\) and
\(K_{ab}=\langle a|K_{\rm tot}|b\rangle\), we evaluated the exact
QFI density from}
{\color{black}
\begin{equation}
f_Q^{(T)}(N)
=
\frac{2}{N}
\sum_{\substack{a,b\\p_a+p_b>0}}
\frac{(p_a-p_b)^2}{p_a+p_b}
|K_{ab}|^2.
\label{eq:S2-QFI-ED}
\end{equation}
}
\textcolor{black}{The smooth ED curves use the same expression in the
common eigenbasis of \(H\) and \(\rho_T\), with
\(p_a=e^{-\beta E_a}/Z\). The \(T=0\) values were evaluated for the
nondegenerate ground states. No Fourier transformation or Gaussian
broadening enters this ED evaluation.
Figure~\ref{fig:S3-ED-TPQ-comparison} compares ED and TPQ for
\(V_z/J_2=0.0,1.0,2.0,3.0\).}

\clearpage

\bibliography{citations}

\end{document}